\documentclass[10pt,journal,final,compsoc]{IEEEtran}
\ifCLASSOPTIONcompsoc
  \usepackage{cite}
\else
  \usepackage{cite}
\fi
\ifCLASSINFOpdf
\else
\fi
\newcommand{\eat}[1]{}
\usepackage{graphicx}
\usepackage{balance}  
\usepackage{booktabs}
\usepackage{tabularx}
\usepackage{amsfonts}
\usepackage{amsmath}
\usepackage{multirow}
\usepackage{algorithm}
\usepackage{algorithmic}
\usepackage{siunitx}

\usepackage{xcolor}
\usepackage{bm}

\DeclareMathOperator{\Tr}{Tr}

\begin{document}
%
\title{Collaborative Learning for Extremely Low Bit Asymmetric Hashing}
%
%
%
%

\author{Yadan~Luo, Zi~Huang, Yang~Li, Fumin~Shen, Yang~Yang and Peng~Cui
\IEEEcompsocitemizethanks{\IEEEcompsocthanksitem Y. Luo, Z. Huang and Y. Li are with School of Information Technology and Electrical Engineering, The University of Queensland, Brisbane 4072, Australia.
E-mail: lyadanluol@gmail.com, huang@itee.uq.edu.au, yang.li@uq.edu.au, .\protect
\IEEEcompsocthanksitem F. Shen and Y. Yang are with the Center for Future Media and the School of Computer Science and Engineering,
University of Electronic Science and Technology of China. E-mail: \{fumin.shen, dlyyang\}@gmail.com.
\IEEEcompsocthanksitem P. Cui is with the Department of Computer Science
and Technology in Tsinghua University, Beijing 100084, China. E-mail: cuip@tsinghua.edu.cn}
\thanks{Manuscript received September 10, 2019; revised February 10, 2020.}}

%
%

\markboth{IEEE TRANSACTIONS ON KNOWLEDGE AND DATA ENGINEERING,~Vol.~14, No.~8, August~2020}%
{Shell \MakeLowercase{\textit{et al.}}: Bare Demo of IEEEtran.cls for Computer Society Journals}
%



\IEEEtitleabstractindextext{%
\begin{abstract}
Hashing techniques are in great demand for a wide range of real-world applications such as image retrieval and network compression. Nevertheless, existing approaches could hardly guarantee a satisfactory performance with the extremely low-bit (e.g., 4-bit) hash codes due to the severe information loss and the shrink of the discrete solution space. In this paper, we propose a novel \textit{Collaborative Learning} strategy that is tailored for generating high-quality low-bit hash codes. The core idea is to jointly distill bit-specific and informative representations for a group of pre-defined code lengths. The learning of short hash codes among the group can benefit from the manifold shared with other long codes, where multiple views from different hash codes provide the supplementary guidance and regularization, making the convergence faster and more stable. To achieve that, an asymmetric hashing framework with two variants of multi-head embedding structures is derived, termed as Multi-head Asymmetric Hashing (MAH), leading to great efficiency of training and querying. Extensive experiments on three benchmark datasets have been conducted to verify the superiority of the proposed MAH, and have shown that the 8-bit hash codes generated by MAH achieve $94.3\%$ of the MAP\footnote{Mean Average Precision (MAP)} score on the CIFAR-10 dataset, which significantly surpasses the performance of the 48-bit codes by the state-of-the-arts in image retrieval tasks.
\end{abstract}

}

\maketitle

\IEEEdisplaynontitleabstractindextext

%
\IEEEpeerreviewmaketitle

\IEEEraisesectionheading{\section{Introduction}\label{sec:introduction}}

\IEEEPARstart{T}{o} diminish the computational and storage cost and to make optimal use of rapidly emerging multimedia data, hashing \cite{survey,survey1,shen1,hu2018angular,DBLP:journals/tmm/ShenYLLTS17,DBLP:journals/tip/ShenZ0SST16,yang,Shen2020TKDE} has attracted much attention from the machine learning community, with a variety of applications on information retrieval \cite{yadan,zero-shot,ruicong,yahui,xingao,zhang2018generative}, person re-identification \cite{reid,reid1}, clustering \cite{zhang2018highly,zhang} and network compression \cite{binary,binary1,binary2}, etc. The goal of hashing is to transform original data structures and semantic affinity into compact binary codes, thereby substantially accelerating the computation with efficient xor operations and saving the storage. 

There are mainly two branches of hashing, i.e., data-independent hashing and data-dependent hashing. For data-independent hashing, such as Locality Sensitive Hashing~\cite{LSH}, there no prior knowledge (e.g., supervised information) about data is available, and hash functions are randomly generated. Nonetheless, expensive storage and computational overhead might be produced since more than $1,024$ bits are usually required to achieve an acceptable performance. To address this problem, research directions turn to data-dependent hashing, which leverages the intrinsic information embedded in the data itself. Roughly, data-dependent hashing can be divided into two categories: unsupervised hashing (e.g., Iterative Quantization (ITQ) \cite{ITQ}), and (semi-)supervised hashing (e.g., Supervised Hashing with Kernels (KSH), Supervised Discrete Hashing (SDH)~\cite{SDH}, Supervised Hashing with Latent Factor Models (LFH)~\cite{LFH}, Column Sampling based Discrete Supervised Hashing (COSDISH)~\cite{COSDISH} and Semi-Supervised Hashing (SSH)~\cite{semihash}). In general, supervised hashing achieves a better performance than unsupervised ones because supervised information (e.g., semantic labels and/or pair-wise data relationship) can help to exploit the intrinsic data property, thereby generating the superior hash codes and hash functions.

With the rapid development of deep learning techniques, deep hashing~\cite{shen1,DPSH,DRSCH,ADSH,hashing1} trained with an end-to-end scheme has been proposed. From the perspective of training strategy, deep hashing could be roughly characterized into two categories: symmetric and asymmetric deep hashing. By assuming both query and database samples share the same distribution, symmetric deep hashing \cite{DPSH,dch,hashnet} leverages a single network to preserve the pair-wise or triplet-wise neighbor affinity, which inevitably results in high complexity, i.e., $\mathcal{O}(n^2)$ or even $\mathcal{O}(n^3)$, where $n$ denotes the number of database points. Asymmetric deep hashing treats query samples and database samples separately based on the asymmetric theory \cite{powerofasy}. Deep asymmetric pair-wise hashing (DAPH) \cite{DAPH} utilizes two distinct mappings to capture variances and discrepancies between the query and database sets, while asymmetric deep supervised hashing (ADSH) \cite{ADSH} learns a hash function only for the query points, thus reducing the training time complexity.



However, most of the existing deep hashing models could hardly guarantee a manageable convergence or performance with the low-bit hash codes (e.g., 4-bit). For example, considering the CIFAR-10 database including images of ten classes, theoretically the minimum number of bits required to represent the full semantics is $4$ (as $4>log_2(10)$), while the 4-bit hash codes generated by ADSH can only achieve $42.7\%$ of the MAP score, which is far from satisfactory. Besides, to reach the performance and storage requirements, it is inevitable and tedious to adjust the default code length and re-tune the network hyper-parameters multiple times in practice. The bit-scalable deep hashing (DRSCH) \cite{DRSCH} is proposed to learn the hash codes of variable length by unequally weighting each bit and then truncating insignificant bits. Nevertheless, DRSCH is still sub-optimal as it directly selects the short hash codes from the long hash codes without considering the latent features, which causes severe information loss and quantization error. Thereby, how to distill critical features from long hash code learning becomes a key question for learning high-quality low-bit hash codes.


Motivated by the aforementioned observations and analyses, in this paper, we propose a novel collaborative learning strategy for extremely low bit (e.g., 4-bit) hashing by simultaneously learning with an auxiliary group of long hash codes with various lengths (e.g., $\{4,8,16\}$-bit). Among the group, the short hash codes can benefit from the manifold shared with other long codes, where multiple views from auxiliary hash codes provide the supplementary guidance and regularization, making the convergence faster and more stable. To achieve this goal, the Multi-head Asymmetric Hashing (MAH) framework is derived, based on the deep asymmetry hashing architecture equipped with the multi-head embedding. As shown in Fig. 1, two variants of multi-head structures (i.e., the flat and cascaded) are explored. The flat one explicitly guides the low-bit embedding branch with multiple supplementary views, while the cascaded one constructs the intermediate layer of low-bit hashing based on the consensus of other hashing learners. The multi-head structure benefits extremely low bit hashing from two perspectives, i.e., 1) it enables the shared intermediate layers to aggregate the gradient flows from all heads and the penultimate layer to select bit-specific representations, adapting the feature distributions to compensate for the information loss; 2) multiple views from different embedding heads on the same training sample provide regularization to extremely low bit hashing. Our main contributions are summarized as follows:  

\begin{itemize}
	\item We propose a novel collaborative learning strategy for deep hashing, aiming to distill knowledge for low-bit hash codes from a group of hashing learners and gain a performance boost. To the best of our knowledge, it is the very first work of introducing model distillation to address the code compression of deep hashing.
	\item Two variants of multi-head structures are derived to efficiently enhance the power of supervision on the shared intermediate layer and benefit bit-specific representation learning. Besides, a group of hash codes with various lengths are jointly learned, which may suit for different platforms without extra inference cost or network re-tuning. 
	\item Experiments on three benchmark datasets demonstrate that the proposed MAH significantly outperforms existing deep hashing methods especially for the low-bit retrieval task and saves up to $20\times$ training time and the storage by $83.33\%$ to $91.67\%$.
\end{itemize}

The paper is organized as follows. Section 2 presents a brief review of the related work, and Section 3 introduces the details of our hashing learning framework. The experimental results, comparisons and component analysis are presented in Section 4, followed by the conclusion and future work in Section 5.

\begin{figure}[t]
	\centering
	\includegraphics[width=0.5\textwidth]{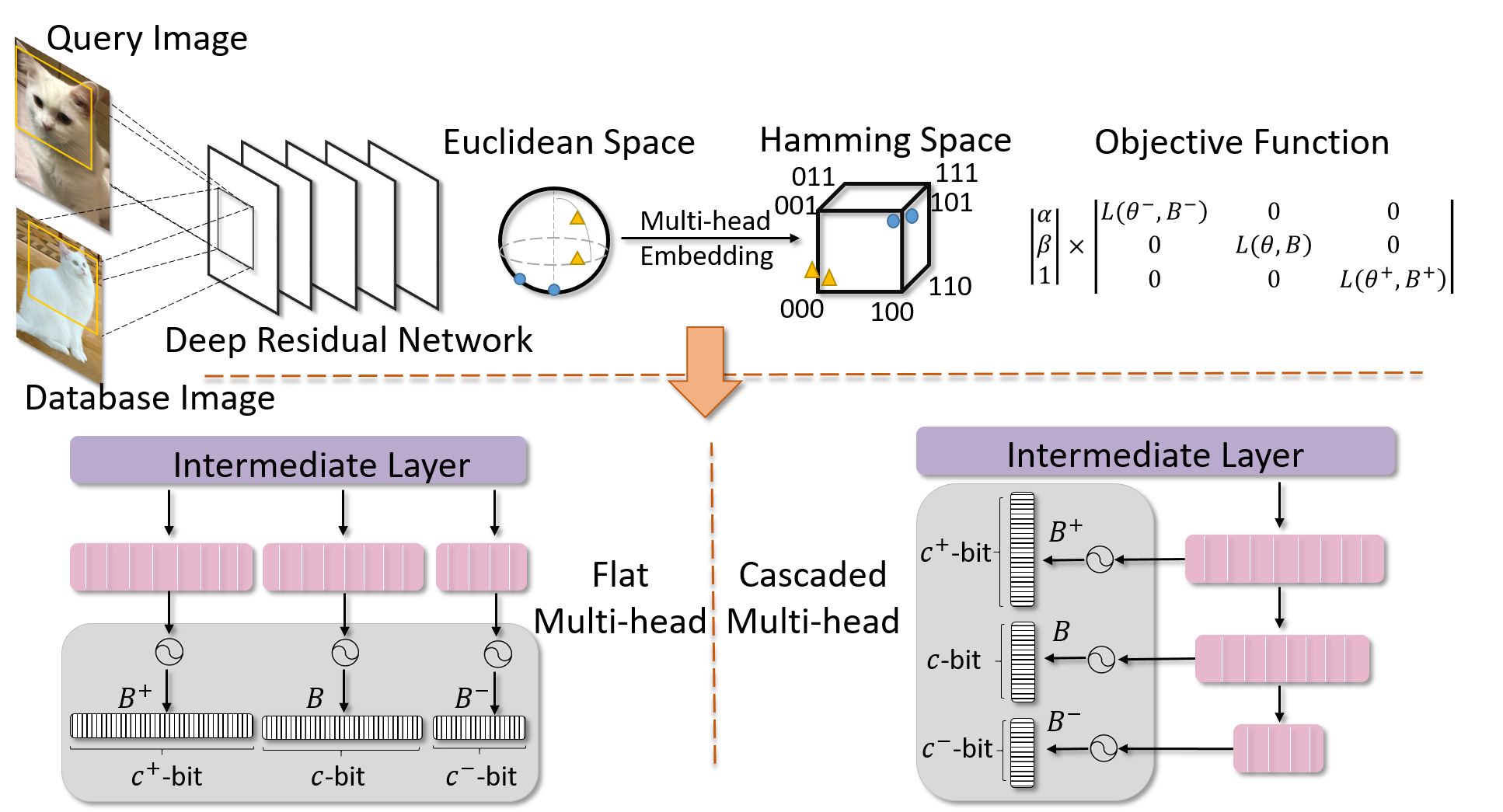}
	\caption{An illustration of the overall pipeline and two variants of multi-head embedding structures. By collaboratively learning with multiple hash learners, the short discrete codes $B^-$ can benefit from the manifold shared with other long codes $B$ and $B^+$, where multiple views from different heads provide the supplementary guidance and regularization, making the convergence faster and more stable.}
	\label{fig:framework}
\end{figure} 

\section{Related Work}
\noindent In this section, we introduce several lines of work that are highly relevant to our research topic: scalable hashing, knowledge distillation and asymmetric hashing. 

\subsection{Scalable Hashing}

Traditional hashing learns hash codes with a default code length (e.g., 64-bit), which highly restricts the flexibility and scalability in practice. For example, low-bit hash codes suit the devices with limited computational resources well, while high-bit hash codes are usually applied in high-performance servers for higher accuracy. Therefore, it is inevitable and tedious for engineers to adjust the default code length and re-tune hyper-parameters of the networks to meet the performance and storage requirements. To address this issue, asymmetric cyclical hashing \cite{asy_cyc} is exploited to measure hash codes with different lengths for query and database images with the weighted Hamming distance. With the development of deep learning techniques, an end-to-end bit-scalable deep hashing framework \cite{DRSCH} is proposed, which learns the hash codes of variable length by unequally weighting each bit and then truncating insignificant bits. Nevertheless, DRSCH is still suboptimal as it ignores the distribution adaptation to fit vertices of the Hamming hypercube and fails to suppress information loss and quantization error.

\subsection{Knowledge Distillation}
For saving computational cost for inferences under various settings, several knowledge distillation strategies have been proposed for classification. While the general knowledge distillation pipeline \cite{hinton} requires two stages, that means, pre-training a large highly regularized model first and then teaching the smaller model, two-way distillation \cite{mutual} leverages an ensemble of students to mutually teach and learn from each other throughout the training process. Nevertheless, using the Kullback Leibler (KL) Divergence to constrain the consensus of prediction and weights of different networks can limit the expression power of the learned representations. As we cannot assume the distributions of the jointly learned embeddings to be either alike or entirely different, it is hard to define a measurement to depict the correlations or dependencies between them. Therefore, we derive two types of multi-head learning, i.e., the flat multi-head structure and the cascaded multi-head structure, which model the implicit and explicit dependencies among the jointly learned discrete codes.

\subsection{Asymmetric Hashing}
Asymmetric hashing could be grouped into two types: dual projection based \cite{DAPH,xingao} and sampling based \cite{ADSH}. Dual projection methods aim to capture the distribution differences between database points and query points by learning two distinctive hash functions so that original data relationships could be well preserved. Rather than learning full pairwise relationships with $\mathcal{O}(n^2)$ complexity or triplet relationships with $\mathcal{O}(n^3)$ among the $n$ points dataset, sampling based methods select $m$ anchors ($m\ll n$) to approximate the query datasets and construct an asymmetric affinity to supervise learning, which significantly reduces the training time complexity to $\mathcal{O}(mn)$.
\section{Methodology}
\color{black}{The overview of the proposed multi-head asymmetric hashing framework is illustrated in Fig. \ref{fig:framework}. We firstly formulate the task of low-bit hashing for similarity-based search, followed by two kinds of multi-head mechanisms, i.e., flat multi-head and cascaded multi-head structures. To enable collaborative learning among different hashing branches, we detail the objective formulation as well as the iterative optimization procedure. To this end, we empirically verify the practicality and scalability of the proposed method by giving the out-of-sample extension and complexity analysis.
\subsection{Notation and Problem Definition}
\color{black}

Without loss of generality, we focus on low bit hashing for image retrieval task with the pair-wise supervision.
We assume that there are $m$ query data points denoted as $\bm{\mathcal{Q}} = \{q_i\}_{i=1}^m$ and $n$ database points denoted as $\bm{\mathcal{X}} = \{x_j\}_{j=1}^n$. Furthermore, pairwise supervised information between $q_i$ and $x_j$ are provided as $\bm{S_{ij}}\in \{-1, +1\}^{m\times n}$, where $S_{ij} = 1$ if $q_i$ and $x_j$ are similar, otherwise $S_{ij} = -1$. The goal of conventional deep hashing is to learn a nonlinear hash function $f(\cdot)$ and generate hash codes for query points $\bm{\widetilde{B}}=\{\tilde{b_i}\}_{i = 1}^m \in \{-1, +1\}^{m\times c}$, and for database points $\bm{B}=\{b_j\}_{j=1}^n \in \{-1, +1\}^{n\times c}$ with minimum information loss, where $c$ is the hash code length. Different from existing deep hashing methods, which learns the fixed-length binary codes, we explore how to jointly optimize hash codes with various lengths. For simplicity, we demonstrate the learning procedure with three different lengths, e.g., $\{c^{-}, c, c^{+}\}$, where $c^{-}$, $c$, $c^{+}$ denote the short-, anchor- and long-length of binary codes respectively. \color{black}\textbf{Our objective} is to learn the shortest database hash codes $\bm{{B}^{-}}\in\{-1, +1\}^{n\times c^-}$ and the optimal query hash function to maximally preserve the knowledge distilled from auxiliary hashing learning of $\bm{B}\in\{-1, +1\}^{n\times c}$ and $\bm{B^+}\in\{-1, +1\}^{n\times c^+}$.\color{black}

\color{black}\subsection{Feature Extraction}
To take advantage of recent advances in deep neural networks, we construct the hash functions on a pre-trained ResNet-50~\cite{resnet} for quickly learning semantics representations from unseen images. It can be integrated with other deep models such as AlexNet~\cite{alexnet} and VGG~\cite{vgg}. Accordingly, the deep feature extraction procedure for query points can be formulated as,
\begin{equation}
\begin{split}
    \bm{\mathcal{H}} &= \{h_i\}_{i=1}^m\in\mathbb{R}^{m\times D},\\
    h_i&= f(q_i) = f(q_i;\Theta_E), 
    \forall i \in \{1,2,...,m\},
\end{split}
\end{equation}
where $f(\cdot)$ denotes the feature extraction network parameterized by $\Theta_E$, $h_i\in\mathbb{R}^{D}$ indicates the $D$-dim latent vector from the last fully conneted layer of ResNet-50 for the query point $q_i$. 
\color{black}
\color{black}\subsection{Collaborative Multi-head Hashing}\color{black}

\color{black}As the shared characteristics are captured by backbone feature extraction, we then propose two mechanisms to learn discrete code from different views. More concretely, two variants of multi-head structures, i.e., the flat multi-head and cascaded multi-head are constructed for collaborative hashing learning based on different assumptions. Without loss of generality, we only showcase the formulation of collaborative learning with three embedding heads, which could be easily extended to the n-head structure. 

\subsubsection{Flat Multi-head Structure}
Inspired by two-way distillation~\cite{mutual}, we first propose the flat multi-head structure that aims to let compact hash codes benefit from the learning with longer codes. As illustrated in the left half of Fig. \ref{fig:framework}, the flat multi-head consists of three separate embedding layers transforming the shared latent vector $h_i$ to different dimensions of space, 
\begin{equation}\label{eq:flat}
\begin{split}
    \tilde{h}_i^- = \mathcal{F}(h_i; \Theta^{-}),
    \tilde{h}_i = \mathcal{F}(h_i; \Theta),
    \tilde{h}_i^+ = \mathcal{F}(h_i; \Theta^{+}),
\end{split}
\end{equation}
where $\mathcal{F}(\cdot; \Theta^-)$, $\mathcal{F}(\cdot; \Theta)$, $\mathcal{F}(\cdot; \Theta^+)$ are linear projections from $D \mapsto c^-, c, c^+$, respectively. The $\tilde{h}_i^-, \tilde{h}_i, \tilde{h}_i^+$ represents the distilled latent embedding from multi-head learning. The flat embedding regime targets at increasing the posterior entropy of low-bit branch, which helps it converge to a more robust and flatter minima with complementary views.   

\subsubsection{Cascaded Multi-head Structure}
Alternatively, we consider the other option of multi-head architecture that can explicitly exploit critical knowledge from high dimension manifolds. As shown in the right half of Fig. \ref{fig:framework}, we derive a cascaded multi-head that directly passes the learned information flow to the next low-dim space, which can be defined as,
\begin{equation}
\begin{split}\label{eq:cascaded}
    \tilde{h}_i^- = \mathcal{F}(\tilde{h}_i; \Theta^{-}),
    \tilde{h}_i = \mathcal{F}(\tilde{h}_i^+; \Theta),
    \tilde{h}_i^+ = \mathcal{F}(h_i; \Theta^{+}).
\end{split}
\end{equation}
Similarly, $\mathcal{F}(\cdot; \Theta^-)$, $\mathcal{F}(\cdot; \Theta)$, $\mathcal{F}(\cdot; \Theta^+)$ are linear projections from  $c \mapsto c^-$, $c^+ \mapsto c$, $D \mapsto c^+$. The cascaded networks learn to partially attend the critical dimensions of features rather than simply learn separate transformations from the full shared features space. By leveraging the consensus from the long-bit learners which convey more original data structure and semantics, the cascaded multi-head gradually approximates the distributions of embedding towards the vertices of the target Hamming hypercube, thereby significantly suppressing the information loss with progressive adaptation.

\subsection{Formulation of Asymmetric Objective Function}
In order to learn asymmetric hash codes that can fully preserve the semantics and affinity from database and query set, the objective function is formulated as two major parts, that is, for maximizing semantics preservation and minimizing quantization error.

To minimize the $L_2$ loss between the pair-wise supervision and the inner product of query-database binary code pairs $\bm{\widetilde{B}}$ and $\bm{B}$, we obtain the following objective $\mathcal{L}_s(c)$ for each branch of hash codes with the length $c$,

\begin{equation}\label{eq:1}
\begin{split}
&\min_{\widetilde{\bm{B}},\bm{B}}\mathcal{L}_s(c) = \sum_{i = 1}^m\sum_{j = 1}^n(\tilde{b_i}^\mathrm{T}b_j - c\bm{S}_{ij})^2 \\
&\text{s.t.}~\tilde{b}_i = \text{sgn}(\tilde{h}_i)\in \{-1, +1\}^c, \forall i \in \{1,2, ..., m\} \\
&b_j \in \{-1, +1\}^c, \forall j \in \{1,2, ..., n\},\\
\end{split}
\end{equation}
where $sgn(\cdot)$ denotes the signum function for binarization. However, there exists an ill-posed gradient problem in Eq. \eqref{eq:1} caused by the non-smooth $sgn(\cdot)$ function, the gradient of which is zero for all nonzero inputs, making the standard back-propagation infeasible. Therefore, we adopt $\text{tanh}(\cdot)$ function to approximate it and apply a further optimization strategy. Besides, we might be only given a set of database points $\bm{\mathcal{X}}$ without the query points $\bm{\mathcal{Q}}$ in practice. In this case, we follow the same strategy of \cite{ADSH} and randomly sample $m$ data points from database to construct the query set. More specifically, we set $\bm{\mathcal{Q}} = \bm{\mathcal{X}}^{\Omega}$, where ${\bm{\Omega}} = \{i_1, i_2, ..., i_m\}\subset \{1,2, ..., m\}$. The sampling strategy not only avoids overfitting on the training dataset and but also reduces training complexity, thus significantly improving its robustness and practicality. Therefore, Eq. \eqref{eq:1} could be rewritten as:
\begin{equation}
\begin{split}
\min_{\Theta_E, \Theta, \bm{B}}\mathcal{L}_s(c)& = \sum_{i\in\Omega}^m\sum_{j = 1}^n(\sigma(\tilde{h}_i)^\mathrm{T}b_j - c\bm{S}_{ij})^2 \\
\text{s.t.} \quad b_j &\in \{-1, +1\}^c, \forall j \in \{1,2, ..., n\},
\end{split}
\label{eq:2}
\end{equation}
where $\sigma$ is the tanh activation function and $\Theta$ is the learned hyper-parameters of $c$-bit branch of multi-head module. As the query points are sampled from query set, we are expected to minimize quantization error between the approximated hash codes $\text{tanh}(\tilde{h})_i$ and the solved hash codes $b_i$, where the objective $\mathcal{L}_q$ for the $c$-bit hashing learning is formulated as follows,
\begin{equation}
\begin{split}
\min_{\Theta_E, \Theta, \bm{B}}\mathcal{L}_q(c)& = \sum_{i\in\Omega}\big(b_j - \sigma(\tilde{h}_i)\big)^2 \\
\text{s.t.} \quad b_j &\in \{-1, +1\}^c, \forall j \in \{1,2, ..., n\},
\end{split}
\label{eq:3}
\end{equation}
To collaboratively optimize all branches from multi-head embedding, we jointly consider the semantic preservation loss and quantization error loss for different lengths $c^{-}$, $c$ and $c^{+}$ altogether. Consequently, the final objective function can be achieved as,
\begin{equation}\label{eq:loss}
\begin{split}
&\min_{\Theta_E, \Theta^-, \Theta, \Theta^+, \bm{B}^{-}, \bm{B}, \bm{B}^{+}}\mathcal{L}=\alpha (\mathcal{L}_s(c^-)+\gamma\mathcal{L}_q(c^-)) \\
&+ \beta (\mathcal{L}_s(c)+\gamma\mathcal{L}_q(c)) + (\mathcal{L}_s(c^+)
+\gamma\mathcal{L}_q(c^+))\\
&\text{s.t.}\quad b_j^- \in \{-1, +1\}^{c^-}, b_j \in \{-1, +1\}^{c},\\
&b_j^{+} \in \{-1, +1\}^{c^+},\forall j \in \{1,2, ..., n\},
\end{split}
\end{equation}
where $\Theta^+$ and $\Theta^-$ denote the hyper-parameters of $c^+$-bit and $c^-$-bit embedding heads respectively. Moreover, $\alpha$ and $\beta$ are the coefficients for balancing collaborative learning. $\gamma$ is the constant parameter that trades off between the semantic preservation loss and quantization loss.

\subsection{Optimization Algorithm}
The optimization of Eq. \eqref{eq:loss} consists of hash function learning for query points and discrete optimization problem for database hash codes.

\subsubsection{Hash Function Learning}
In order to learn the parameters of the shared embedding module and multi-head structure, we use back-propagation (BP) for gradient calculation. Specifically, we sample a mini-batch of the query points, then update the parameter groups based on the sampled data. Accordingly, we can leverage gradient descent to achieve the optimal value of parameter groups,

\begin{equation}
\begin{split}
\Theta_E^*&\gets\Theta_E-\eta\cdot\nabla_{\Theta_E}\frac{1}{m}\mathcal{L},\\
{\Theta^+}^*&\gets\Theta^+-\eta\cdot\alpha\nabla_{\Theta^+}\frac{1}{m}(\mathcal{L}_s(c^+)+\gamma \mathcal{L}_q(c^+)),\\
{\Theta}^*&\gets\Theta-\eta\cdot\beta\nabla_{\Theta}\frac{1}{m}(\mathcal{L}_s(c)+\gamma \mathcal{L}_q(c)),\\
{\Theta^-}^*&\gets\Theta^--\eta\cdot\nabla_{\Theta^-}\frac{1}{m}(\mathcal{L}_s(c^-)+\gamma \mathcal{L}_q(c^-)),
\end{split}
\label{eq:theta}
\end{equation}
where $m$ is the batch size, $\eta$ is the learning rate. We use the chain rule to compute gradient flows to update $\Theta_E$, $\Theta^+$, $\Theta$, and $\Theta^-$ asynchronously.

\subsubsection{Discrete Hash Codes Solution}
In this subsection, we aim to solve the subproblem of optimizing binary codes for database set with all network parameters fixed. Due to the discrete constraints of hash codes, it is NP hard to achieve a closed-form solution directly. \color{black}Inspired by the discrete cyclic coordinate descent (DCC)~\cite{SDH}, the binary codes could be learned bit by bit iteratively. Firstly we target at optimizing $\bm{B}$ and fix all other variables, and rewrite the Eq. \eqref{eq:loss} as follows,

\color{black}
\begin{equation}\label{eq:DCC}
\begin{split}
\min_{\bm{B}} \mathcal{L}(c) = \|\bm{\widetilde{\mathcal{H}}}\bm{B}^\mathrm{T}& - c\bm{S}\|_F^2 + \gamma \|\bm{B}^{\Omega}-\bm{\widetilde{\mathcal{H}}}\|_F^2,\\
= \|\bm{\widetilde{\mathcal{H}}}\bm{B}^\mathrm{T}\|_F^2 -2c\Tr&(\bm{B}^\mathrm{T}\bm{S}^\mathrm{T}\bm{\widetilde{\mathcal{H}}})-2\gamma\Tr(\bm{B}^{\Omega}\bm{\widetilde{\mathcal{H}}}^\mathrm{T}) + const,\\
\text{s.t.}~\bm{B}\in\{-1, +1&\}^{n\times c},
\end{split}
\end{equation}
where $\|\cdot\|_F^2$ denotes the Frobenius norm, $\bm{\widetilde{\mathcal{H}}}\in \mathbb{R}^{m\times c}$ is the output from $c$-bit branch of multi-head embedding, $\bm{B}^{\Omega}\in\{-1, +1\}^{m\times c}$ denotes the binary codes for the database points whose rows are selected by $\bm{\Omega}$, and $const$ is the constant. For simplicity, we reshape the $\bm{\widetilde{\mathcal{H}}}\in\mathbb{R}^{m\times c}$ to the matrix $\bm{\widehat{\mathcal{H}}}\in\mathbb{R}^{n\times c} = \{\hat{h}_i\}_{i=1}^n$, where $\hat{h}_i = \tilde{h}_i ~\text{if}~i\in\bm{\Omega}$ otherwise $\hat{h}_i=0$. Besides, we define a binary mask matrix $\bm{M_{\Omega}}\in\{-1, +1\}^{n\times c}$, where the $i$-th row is assigned $1$ if $i\in\bm{\Omega}$ otherwise assigned with $0$. Hence the problem \eqref{eq:DCC} can be rewritten as,

\begin{equation}
\begin{split}
\min_{\bm{B}}\mathcal{L}(c)&= \|\bm{\widetilde{\mathcal{H}}}\bm{B}^\mathrm{T}\|_F^2-2\Tr(c\bm{B}^\mathrm{T}\bm{S}^\mathrm{T}\bm{\widetilde{\mathcal{H}}}+\gamma\bm{B}\odot\bm{M_{\Omega}}\bm{\widehat{\mathcal{H}}}^\mathrm{T}),\\
&= \|\bm{\widetilde{\mathcal{H}}}\bm{B}^\mathrm{T}\|_F^2 + \Tr(\bm{B}^\mathrm{T}\bm{E})+ const,\\
\text{s.t.}\,\bm{E}=-2(&c\bm{S}^\mathrm{T}\bm{\widetilde{\mathcal{H}}}+\gamma \bm{M_\Omega}^\mathrm{T}\odot\bm{\widehat{\mathcal{H}}}), \quad \bm{B}\in\{-1, +1\}^{n\times c}.
\end{split}
\label{eq:8}
\end{equation}
As $\bm{B}$ is discrete and non-convex, we choose to learn the binary codes $\bm{B}$ by the discrete cyclic coordinate descent (DCC) method. In other words, we learn $\bm{B}$ bit by bit. Let $\bm{B}_{*k}$ denote the $k$-th column of $\bm{B}$, and $\bm{B}'$ denote the matrix of $\bm{B}$ excluding the $k$-th column. Similarly, we let $\bm{E}_{*k}$ denote the $k$-th column of $\bm{E}$, and $\bm{E}'$ denote the matrix of $\bm{E}$ excluding the $k$-th column. Let $\bm{\widetilde{\mathcal{H}}}_{*k}$ denote the $k$-th column of $\bm{\widetilde{\mathcal{H}}}$, and $\bm{\widetilde{\mathcal{H}}}'$ denote the matrix of $\bm{\widetilde{\mathcal{H}}}$ excluding the $k$-th column. To optimize $\bm{B}_{*k}$, we can calculate the following objective function,

\begin{equation}
\begin{split}
\min_{\bm{B}}\mathcal{L}(c) =& \|\bm{\widetilde{\mathcal{H}}}\bm{B}^\mathrm{T}\|_F^2 + \Tr(\bm{B}^\mathrm{T}\bm{E}),\\
=& \Tr(\bm{B}_{*k}[2\bm{\widetilde{\mathcal{H}}}_{*k}^\mathrm{T}\bm{\widetilde{\mathcal{H}}}'\bm{B}'+\bm{E}_{*k}]),\\
\text{s.t.} \quad &\bm{B}_{*k} \in \{-1, +1\}^n.
\end{split}
\end{equation}
Consequently, the optimal solution of the $k$-th column of $\bm{B}$ could be achieved as,
\begin{equation}
\begin{split}
\bm{B}_{*k} = -\text{sgn}(2\bm{\widetilde{\mathcal{H}}}_{*k}^\mathrm{T}\bm{\widetilde{\mathcal{H}}}'\bm{B}'+\bm{E}_{*k}).
\end{split}
\label{eq:B}
\end{equation}
In the similar way, the solution of $B^+$ and $B^{-}$ can be obtained with all the other variables fixed through iterative update bit by bit,

\begin{equation}
\begin{split}\label{eq:finalB}
\bm{B}^{+}_{*k} = -\text{sgn}(2{\bm{\widetilde{\mathcal{H}}}_{*k}^{+\mathrm{T}}}\bm{\widetilde{\mathcal{H}}}^{+'}{\bm{B}^{+'}}+\bm{E}_{*k}^+),\\
\bm{B}^{-}_{*k} = -\text{sgn}(2{\bm{\widetilde{\mathcal{H}}}_{*k}^{-\mathrm{T}}}\bm{\widetilde{\mathcal{H}}}^{-'}{\bm{B}^{-'}}+\bm{E}_{*k}^-),
\end{split}
\end{equation}
where $\bm{E}^+=-2(c^+\bm{S}^\mathrm{T}\bm{\widetilde{\mathcal{H}}}^++\gamma \bm{M_\Omega}^{+\mathrm{T}}\odot\bm{\widehat{\mathcal{H}}}^+)$ and $\bm{E}^-=-2(c^-\bm{S}^\mathrm{T}\bm{\widetilde{\mathcal{H}}}^-+\gamma \bm{M_\Omega}^{-\mathrm{T}}\odot\bm{\widehat{\mathcal{H}}}^-)$. The complete training procedure for the proposed MAH is described in Algorithm 1. 

\begin{algorithm}[!thb]
	\begin{algorithmic}[1]
		\renewcommand{\algorithmicrequire}{\textbf{Input:}}
		\renewcommand{\algorithmicensure}{\textbf{Output:}}
		\REQUIRE $\bm{\mathcal{X}}=\{x_i\}_{i=1}^n$: $n$ data points; $\bm{S}\in\{-1,+1\}^{n\times n}$: supervised similarity matrix; $\{c^{-}, c, c^{+}\}$: binary code lengths.
		\ENSURE ${\bm{B}^-, \bm{B}, \bm{B}^+}$: Binary codes for database; $\Theta^-, \Theta, \Theta^+$: Parameters of the multi-head embedding; $\Theta_E$: Parameter of the shared feature learning;
		\STATE Initialize parameters and binary codes, batch size $M$, the number of epochs $T$; the number of sampling query sets $K$;
		\FOR{$j=1$ to $K$}
		\STATE Randomly sample the index $\bm{\Omega}$ and set $\bm{S} = \bm{S}^{\Omega}$, $\bm{\mathcal{Q}} = \bm{\mathcal{X}}^{\Omega}$;
		\FOR{$t=1$ to $T$}
		\FOR{$i=1$ to $\frac{m}{M}$}
		\STATE Construct a mini-batch and calculate the shared feature $h_i$ and multi-head embedding $\tilde{h_i}^-, \tilde{h_i}, \tilde{h_i}^+$ for each data point in the mini-batch by forward propagation.
		\STATE Optimize the parameters $\{\Theta_E, \Theta^-, \Theta, \Theta^+\}$ using back propagation according to Eq. \eqref{eq:theta}.
		\ENDFOR
		\FOR{$k=1$ to $c^+$}
		\STATE{Update the $\bm{B}^+_{*k}$ according to Eq. \eqref{eq:finalB};}
		\ENDFOR
		\FOR{$k=1$ to $c$}
		\STATE{Update the $\bm{B}_{*k}$ according to Eq. \eqref{eq:B}}
		\ENDFOR
		\FOR{$k=1$ to $c^-$}
		\STATE{Update the $\bm{B}^-_{*k}$ according to Eq. \eqref{eq:finalB};}
		\ENDFOR
		\ENDFOR
		\ENDFOR
	\end{algorithmic}
	\caption{Pseudocode of Optimizing the Proposed MAH}
	\label{alg:1}
\end{algorithm}

\subsection{Out-of-Sample Extension}
After training the MAH, the learned deep neural networks can be applied for generating compact binary codes for query points including unseen query points (e.g., $x_u$) during training. Specifically, we can use the following equation to generate the compact hash codes for $x_u$,

\begin{equation}
b_u^- = \text{sgn}\Big(\mathcal{F}\big(f(x_u; \Theta_E); \Theta^-\big)\Big).
\end{equation}

\color{black}
\subsection{Complexity Analysis}
For each epoch, the time cost is analyzed as follows. The computation of gradient of $\{\tilde{h}^-, \tilde{h}, \tilde{h}^+\}$ in Eq. \eqref{eq:flat} or Eq. \eqref{eq:cascaded} is $\mathcal{O}(nm(c^-+c+c^+)+2m(c^-+c+c^+))$. To apply DCC algorithm for updating $\{B^-,B,B^+\}$, $\{E^-, E, E^+\}$ in Eq. \eqref{eq:8} to be calculated with the cost of $\mathcal{O}(3nm^2 + n(c^-+c+c^+))$. Regarding to the optimization of the sub-problem in Eq. \eqref{eq:B} and \eqref{eq:finalB}, the time cost is $\mathcal{O}([(c^--1)^2+(c-1)^2+(c^+-1)^2]n + (c^+ + c + c^- -3)m^2)$. In practice, $m, c^-, c, c^+$ are much smaller than the database size $n$. Hence, the overall computational complexity of the proposed algorithm MAH is $\mathcal{O}(n)$.

\section{Experiments}
\color{black}
We conduct experiments on three commonly-used benchmarks to compare our method with the state-of-the-art methods. The images in the datasets are in a wide spectrum of image types and label types, including tiny objects from the single-labeled \textit{CIFAR-10}, web images from the \textit{NUS-WIDE} and the \textit{MIRFlickr} with multi-label annotations. The datasets and evaluation protocols are summarized as follows.
\color{black}

\begin{table*}
	\caption{The MAP score (\%) of the proposed MAH and the baselines with short code lengths performing on three large-scale datasets. MAH-1 and MAH-2 denote the proposed method equipped with flat multi-head and cascaded multi-head structure, respectively. \newline $^*$ indicates a re-implementation with the official source code and the default hyper-parameters.}
	\resizebox{1\textwidth}{!}{%
		\begin{tabular}{l|ccccc|ccccc|ccccc}
			\toprule
			\multirow{1}{*}{Method} &
			\multicolumn{5}{c}{CIFAR-10} &
			\multicolumn{5}{c}{NUS-WIDE} &
			\multicolumn{5}{c}{MIRFLickr} \\
			& {4 bits} & {6 bits} & {8 bits} & {10 bits} & {12 bits} & {4 bits}  & {6 bits} & {8 bits} & {10 bits} & {12 bits}& {4 bits} & {6 bits} & {8 bits} & {10 bits} & {12 bits}\\
			\midrule
			LSH &11.28  &11.71 &12.64  &12.13  &12.09 &45.65 &45.89 &43.76 &44.38 &44.11 &54.26 &56.73 &55.16 &57.16 &56.49 \\
			ITQ &18.22  &18.00 &19.39  &19.76  &20.16  &62.56 &67.07 &69.06 &70.39 &70.82  &71.56 &71.69 &71.46 &72.54 &73.13\\
			\midrule
			SDH &34.79 &43.23 &44.83 &52.49 &56.01 &62.92 &65.98 &68.53 &73.33 &71.73 &76.09 &78.14 &80.15 &80.38 &81.58\\
			KSH &34.81 &40.98  &44.96  &46.23  &47.22 &59.19 &59.76 &59.85 &59.40 &59.66 &66.29 &66.06 &66.36 &68.99 &68.33\\
			LFH &30.06 &15.34  &36.81  &37.93  &45.14 &58.48 &68.90 &68.08 &72.81 &72.25 &68.71 &74.12 &81.34 &81.55 &84.07\\
			COSDISH &59.24 &65.55  &70.31  &73.99  &74.81 &58.49 &66.16 &70.62 &73.57 &73.62 &68.73 &80.31 &79.28 &77.83 &82.06 \\
			\midrule
			DPSH$^*$ &38.59  &50.09 &54.12  &61.23 &65.34 &52.63 &57.68 &64.43 &67.14 &70.15 &39.96 &52.19 &55.73 &59.83 &65.86\\
			\midrule
			DRSCH &40.23  &49.87  &56.92  &60.15  &67.54 &48.07 &49.53 &52.42 &53.60 &55.35 &42.35 &51.92 &58.76 &63.51&67.21\\
			\midrule
			ADSH &42.69 &50.06 &87.30 &92.48 &92.84 &73.31 &74.63 &76.85 &78.68 &79.13 &72.68 &82.13 &83.61 &85.58 &86.54\\
			DAPH &53.48 &61.88 &66.48 &73.28 &75.69  &53.26 &56.70 &64.76 &68.30 &71.67 &70.25 &78.30 &80.12 &85.43 &87.12 \\
			\midrule
			\textbf{MAH-1} &47.59 &\textbf{81.97} &\textbf{93.39} &\textbf{93.35} &\textbf{95.03} &70.95 &\textbf{75.37} &76.79 &\textbf{78.87} &\textbf{79.52} &\textbf{82.64} &\textbf{85.39} &\textbf{86.90 }&\textbf{87.82}  &\textbf{89.20} \\
			\textbf{MAH-2} &\textbf{74.60} &\textbf{89.50} &\textbf{94.29} &\textbf{94.89} &\textbf{95.37} &\textbf{76.47 }&\textbf{76.85} &\textbf{79.47} &\textbf{82.15} &\textbf{84.65} &\textbf{82.75} &\textbf{86.09} &\textbf{86.76} &\textbf{86.81} &\textbf{89.39}\\
			\bottomrule
		\end{tabular}%
	}
	\label{tab:1}
\end{table*}

\subsection{Datasets} 
\textbf{CIFAR-10}~\cite{cifar} labeled a subset of 80 million tiny images, which consists of 60,000 32$\times$32 color images in 10 classes, with 6,000 images per class. \\
\textbf{NUS-WIDE}~\cite{nuswide} is a web image dataset containing 269,648 images from Flickr, where 81 semantic concepts are provided for evaluation. We eliminate all empty images and use the rest 195,834 images from the 21 most frequent concepts, where each concept consists of at least 5,000 images. \\
\textbf{MIRFlickr}~\cite{huiskes08} is a collection of 25,000 images from Flickr, where each instance is manually annotated with at least one of 38 labels. 

Each dataset is randomly split into a query set with 1,000 samples and a database set with the remaining samples for evaluation. For the single-labeled datasets, if two samples have the same class label, they are considered to be semantically similar, and dissimilar otherwise. For multi-labeled datasets, if two samples share at least one semantic label, they are considered to be semantically similar.

\subsection{Evaluation Protocols}
The Hamming ranking is used as the search protocol to evaluate our proposed approach, and two indicators are reported. 

1) Mean Average Precision (\textbf{MAP}): The average precision (AP) is defined as,
\begin{equation}
\text{AP} = \frac{1}{R}\sum_{r=1}^n\text{Precision}(r)\delta(r),
\end{equation}
where $R$ is the number of ground-truth neighbors of the query in a database, $n$ is the number of samples in the database. $\text{Precision(r)}$ denotes the precision of the top $r$ retrieved entities, $\delta(r)=1$ if the $r$-th retrieved entity is a ground-truth neighbor and $\delta(r)=0$ otherwise. For a query set whose size is $m$, the MAP is defined as the mean of the average precision scores for all the queries in the query set,
\begin{equation}
\text{MAP} = \frac{1}{m}\sum_{i=1}^m \text{AP}_i.
\end{equation}

2) Top5000-Precision (\textbf{Precision@5000}): the Top5000-precision curve reflects the change of precision with respect to the number of top-ranked $5000$ instances returned to the users, which is expressive for retrieval.

\subsection{Baselines}
To evaluate the proposed MAH, we select a number of representative hashing methods as baselines for comparison, including data-independent hashing method \textbf{LSH}~\cite{LSH}, unsupervised hashing method \textbf{ITQ}~\cite{ITQ}, four supervised but non-deep supervised hashing methods \textbf{KSH} \cite{KSH}, \textbf{SDH}~\cite{SDH}, \textbf{LFH}~\cite{LFH}, \textbf{COSDISH}~\cite{COSDISH}, four deep supervised hashing methods including \textbf{DPSH}~\cite{DPSH}, \textbf{DRSCH}~\cite{DRSCH}, \textbf{ADSH}~\cite{ADSH} and \textbf{DAPH}~\cite{DAPH}. For non-deep hashing methods, we utilize 4,096-dim deep features which are extracted from the pre-trained ResNet50 model on ImageNet dataset for fair comparison. KSH and SDH are kernel based methods, for which we randomly select 1,000 data points as anchors to construct the kernels by following the suggestion of the authors. 

\begin{figure*}[htb!]
	\centering
	\includegraphics[width=0.32\textwidth]{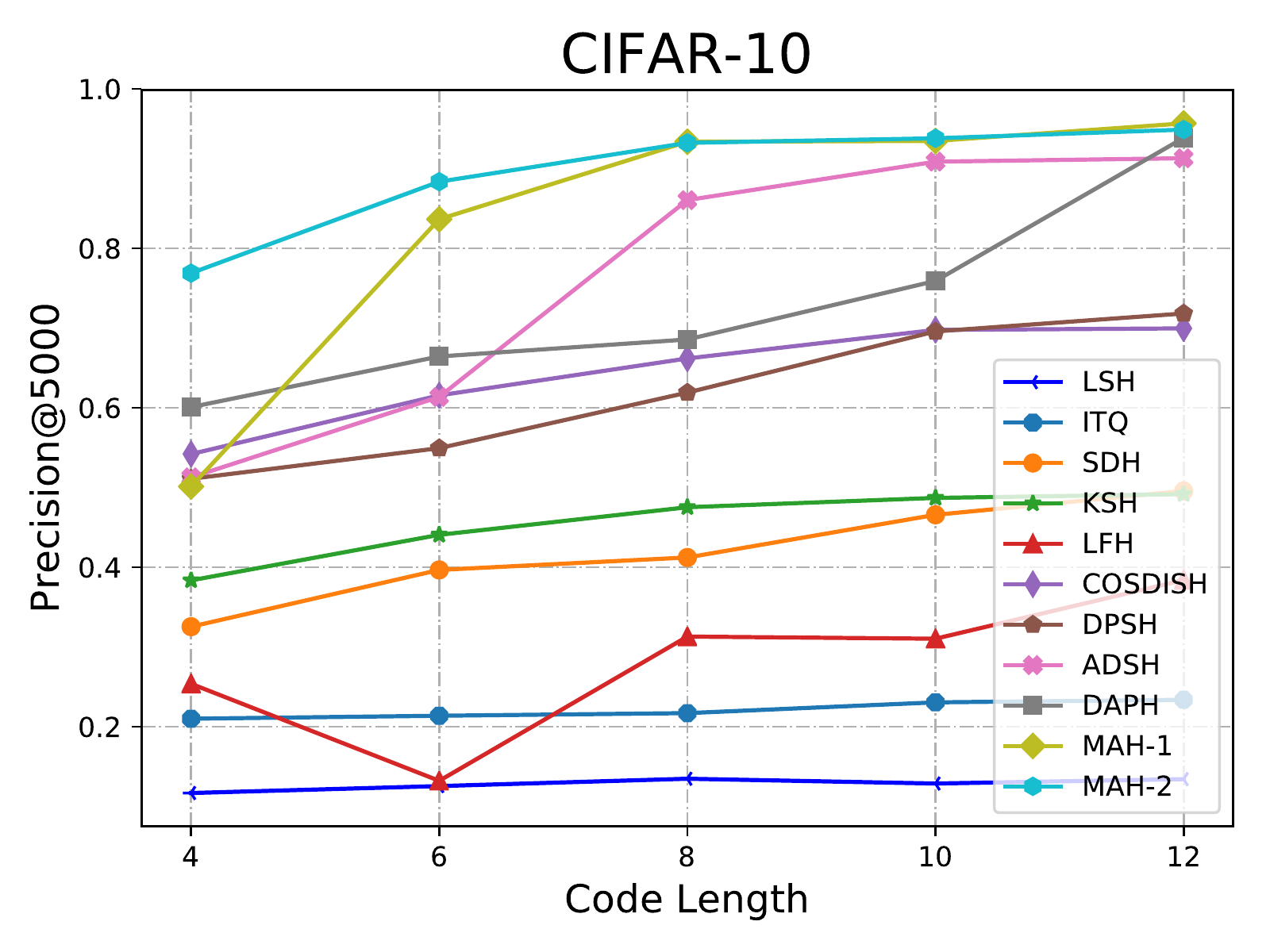}
	\includegraphics[width=0.32\textwidth]{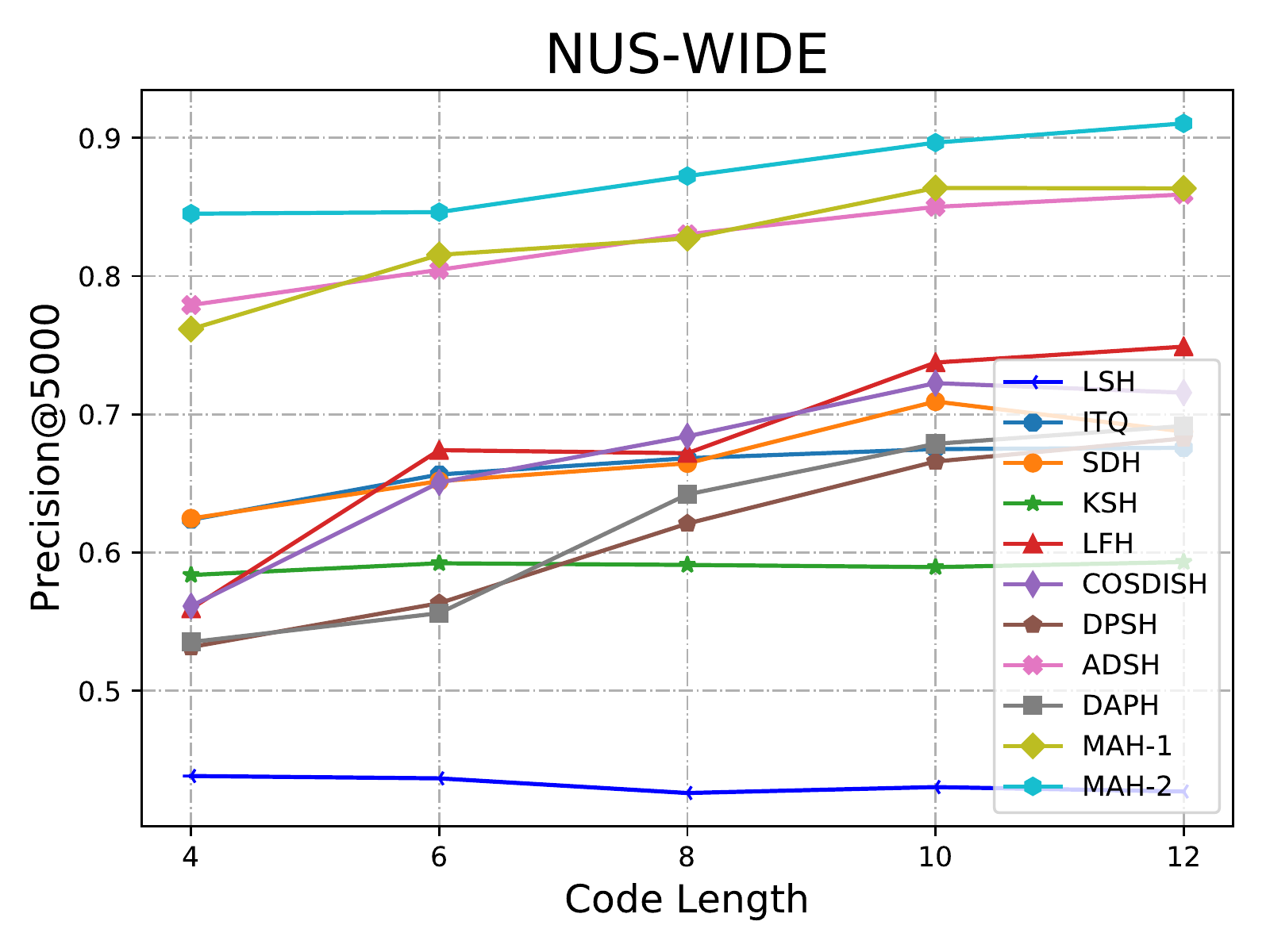}
	\includegraphics[width=0.32\textwidth]{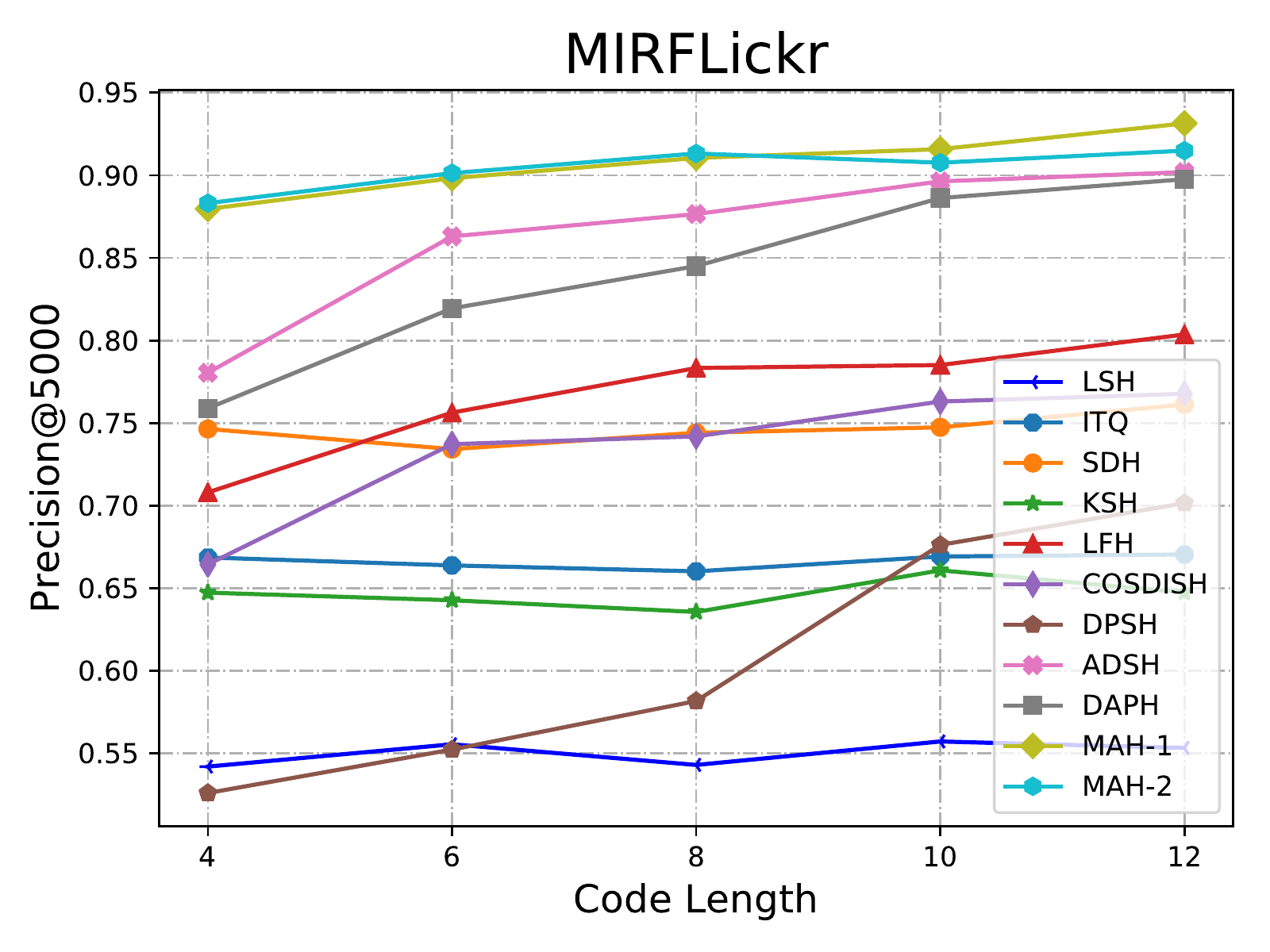}
	\caption{The Precision@5000 curves of the proposed MAH and other baselines on three datasets with respect to different code lengths.}
	\label{fig:precision}
\end{figure*}
\subsection{Implementation Details}
Our algorithm is implemented with Pytorch~\cite{pytorch} \color{black}  and the source code is made available on Github\footnote{\color{black}https://github.com/Luoyadan/MAH-Pytorch} for reference. \color{black}Training is conducted on a server with two Tesla K40c GPUs with 12GB memory. For a fair comparison, we employ the deep residual network (ResNet-50) architecture as the backbone to extract features with dimension $D=1000$. During the training, we use the stochastic gradient descent with the momentum to 0.9 and weight decay to $5\times 10^{-4}$. The batch size is set to 64 and the learning rate is set to $10^{-3}$.
\begin{figure*}[htb!]
	\centering
	\includegraphics[width=0.242\textwidth]{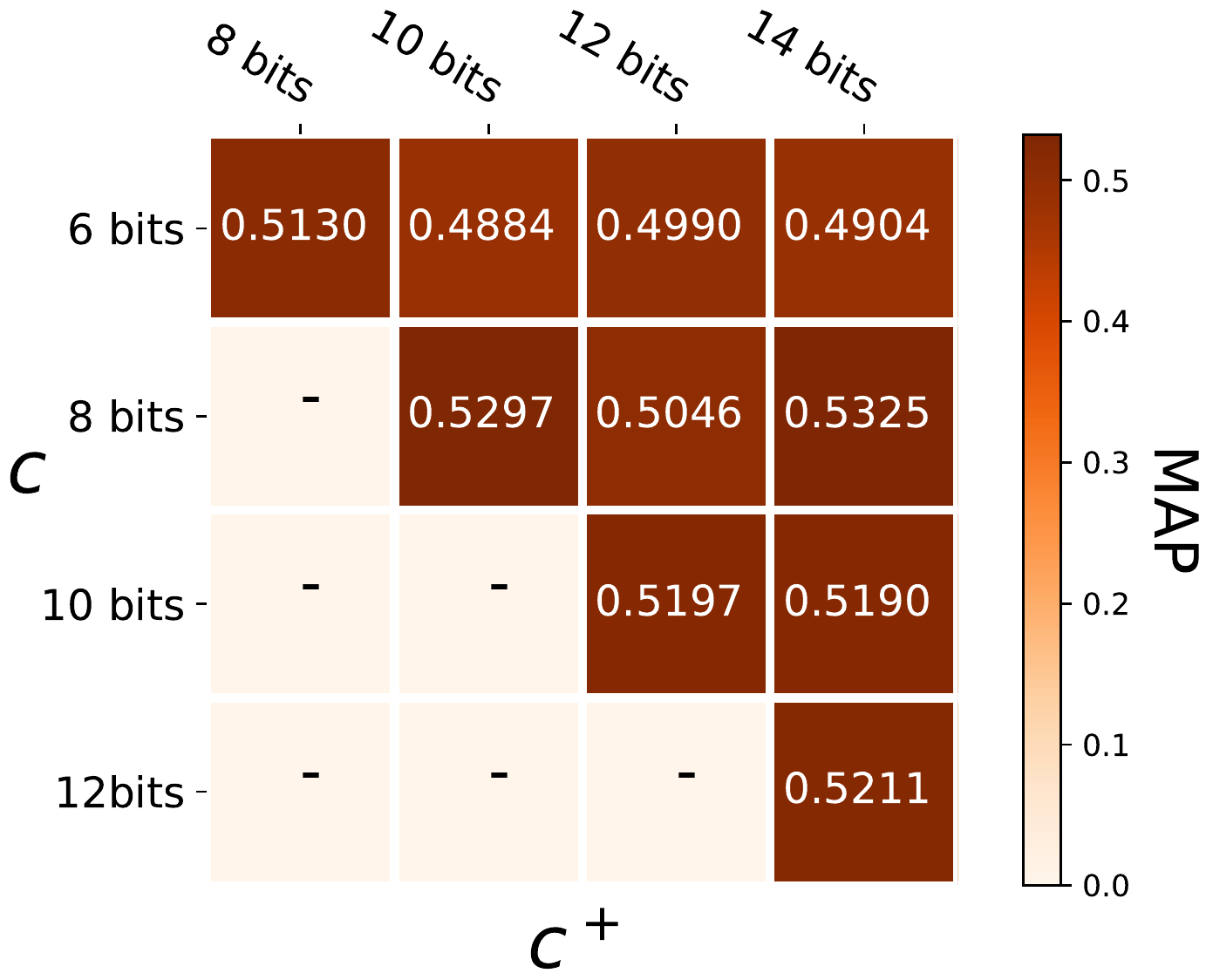}
	\includegraphics[width=0.242\textwidth]{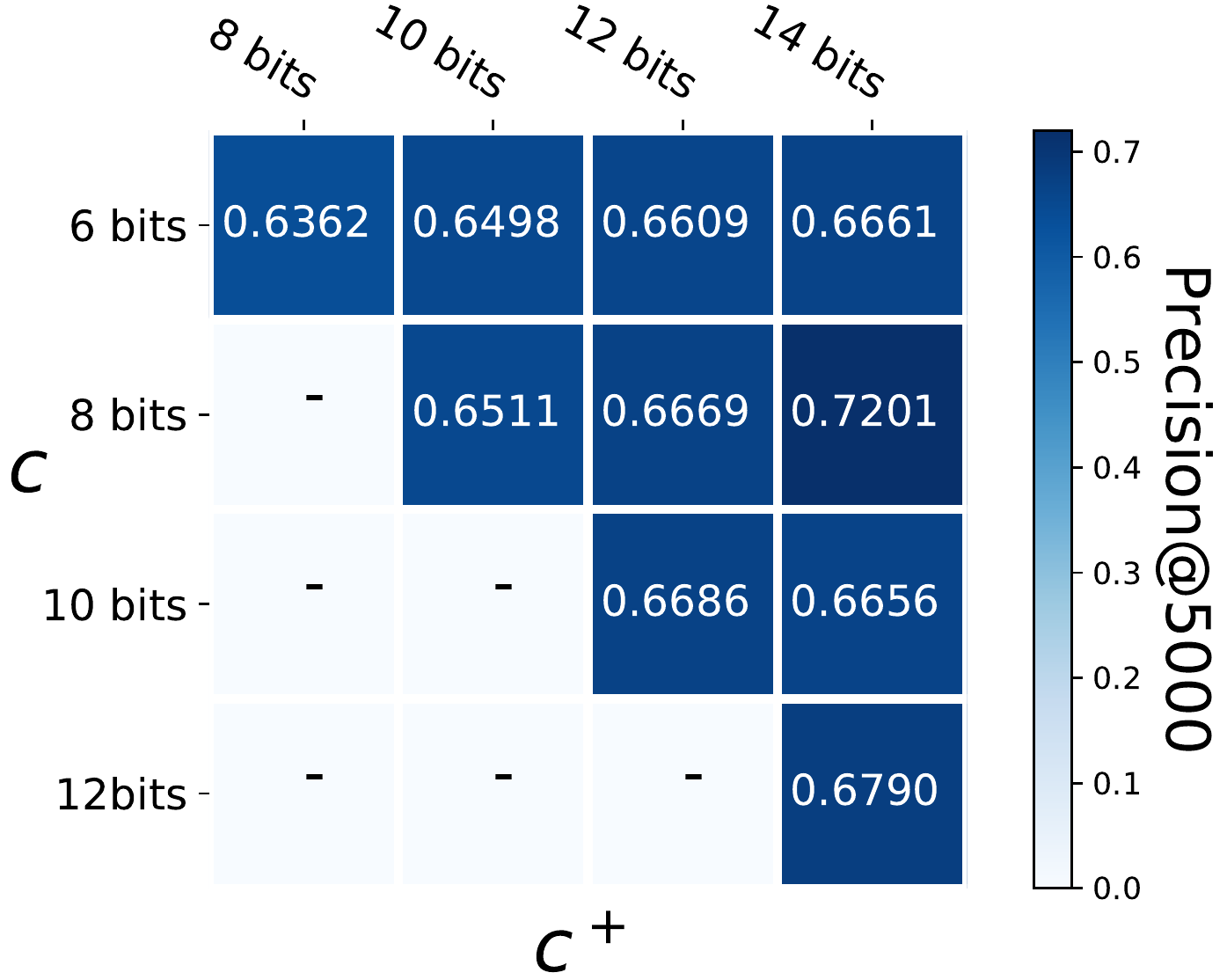}
	\includegraphics[width=0.242\textwidth]{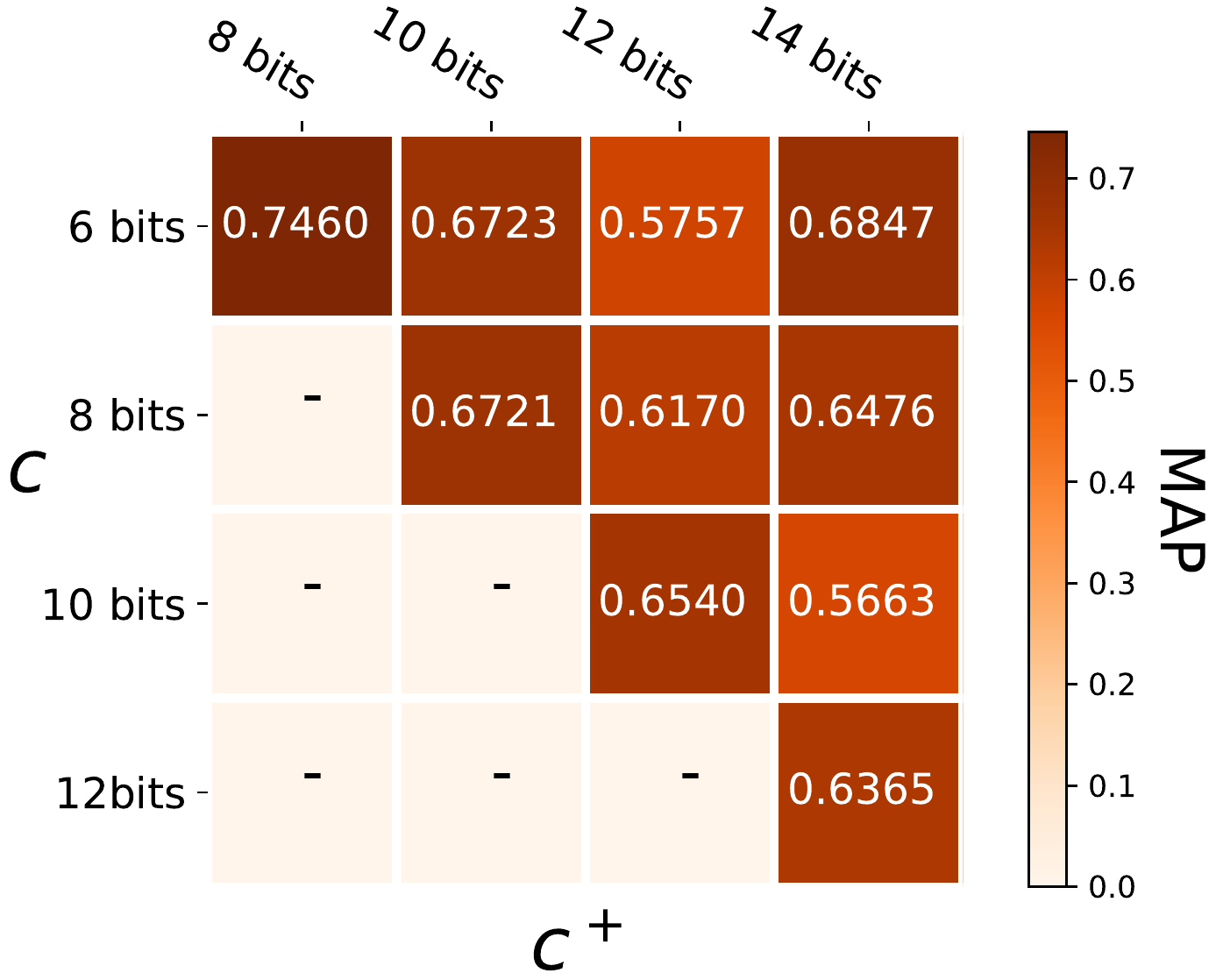}
	\includegraphics[width=0.242\textwidth]{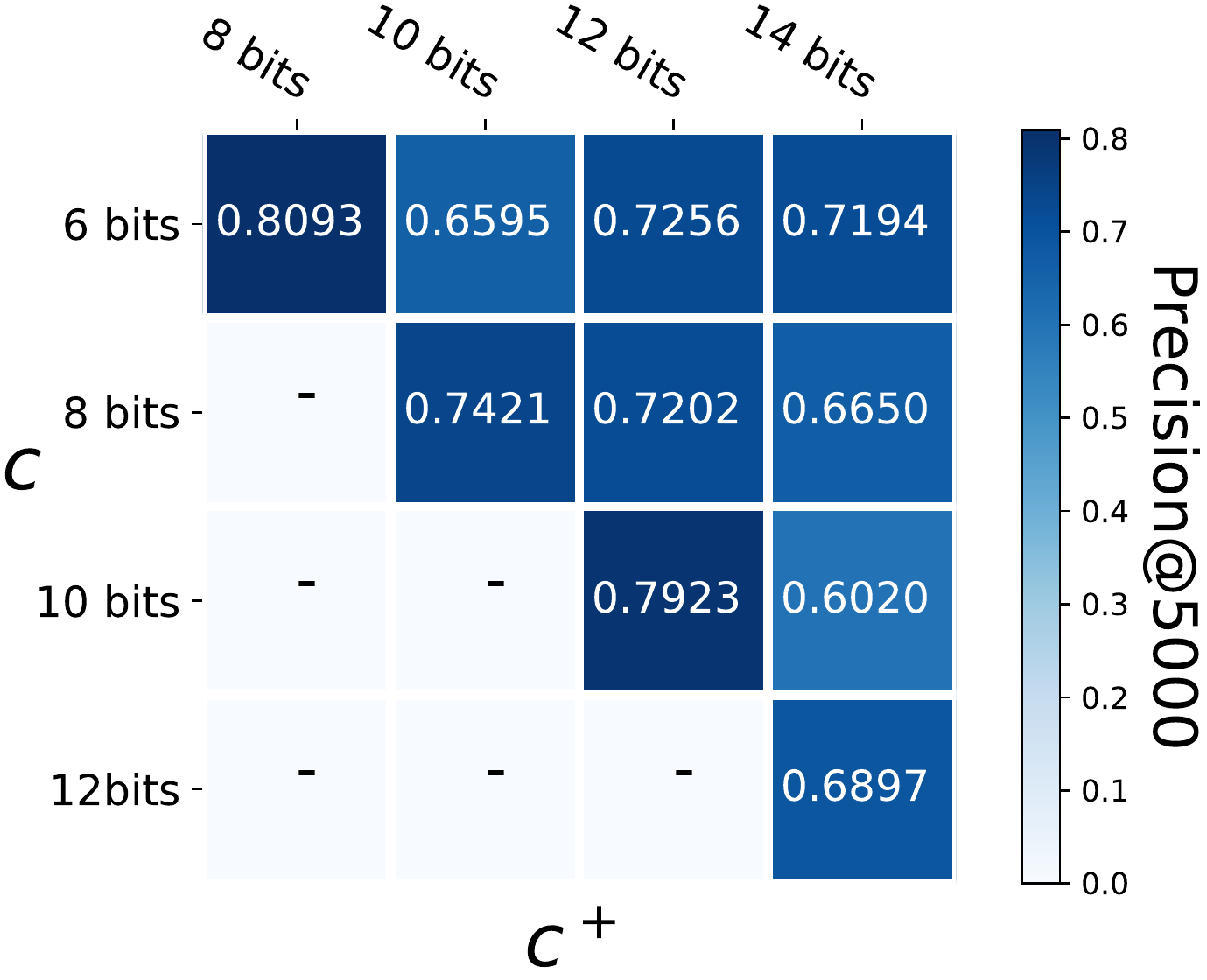}
	\caption{Ablation study on the large-scale multi-head strides ($6$, $8$, $10$, $12$, $14$). The performance of $c^-=4$ bit hash codes learned by MAH with the flat multi-head (\textit{Left}) and cascaded multi-head (\textit{Right}) on CIFAR-10 dataset. The loss coefficients  $\alpha$ and $\beta$ are fixed at 4 and 2, respectively.}
	\label{fig:distiller}
\end{figure*}
\subsection{Comparisons with the State-of-the-art Algorithms}
Table \ref{tab:1} reports the MAP scores of the proposed method and the compared baselines using various code lengths, and Fig. \ref{fig:precision} displays the Precision@5000 curve with 4, 6, 8, 10, 12 bits. Empirically, the loss coefficients $\alpha$ and $\beta$ of MAH are fixed at 6 and 2 respectively and the code length groups are set as $\{j, j+2, j+4\}$, $j \in\{4, 6, 8, 10, 12\}$. From the Table \ref{tab:1} and Fig. \ref{fig:precision}, we can observe that,
\begin{itemize}
	\item The proposed \textbf{MAH} consistently outperforms all baselines on three datasets in all cases, which verifies its validity and effectiveness. Especially when embedding extremely low-bit (e.g., 4-bit) hash codes, MAH achieves at least $39.5\%$ (CIFAR-10), $4.3\%$ (NUS-WIDE), $13.9\%$ (MIRFLickr) higher performance compared with other deep hashing approaches. 
	\item Data-independent and unsupervised hashing methods, i.e., \textbf{LSH} and \textbf{ITQ} achieve a relatively lower performance compared with all supervised methods on the single-label dataset (CIFAR-10), while they gain competitive scores on multi-label datasets. The major reason we infer is the pre-defined similarity measurement, which highly restrains the entropy of the ground-truths. For multi-label datasets, if two samples share at least one semantic label, they are considered to be semantically similar. However, it makes pair-wise relationship vague and fuzzy, hence easier for unsupervised method to predict. 
	\item Supervised yet non-deep methods \textbf{SDH}, \textbf{KSH}, \textbf{LFH}, \textbf{COSDISH}, generally achieve a stable increase on MAP as the hash code length goes up. Notably, supervised methods that adopt discrete optimization, i.e., SDH, COSDISH perform relatively better compared with relaxed hashing approaches.
	\item \textbf{DPSH}, \textbf{DRSCH}, \textbf{ADSH}, \textbf{DAPH} and \textbf{MAH} are all trained in an end-to-end scheme. Compared with the classic DPSH framework, DRSCH applies a bit-wise weight layer to truncate insignificant bits, which benefits the low bit learning and increases MAP by $1.64\%$. DAPH and ADSH are deep asymmetric hashing, which loss 21.12\% and 31.91\% on MAP in comparison with MAH as their fix-length embedding has no guarantee of the convergence to a global minimum.  
	\item The underlying principle is that the collaborative learning strategy adopted by \textbf{MAH} achieves the consensus of multiple views from embedding heads on the same training sample. The consensus provides the supplementary information as well as the regularization to each embedding head, therefore enhancing the generalization and robustness. Besides, the intermediate-level representations shared with back-propagation rescaling aggregate the gradient flows from all heads, which not only reduces training computational complexity, but also facilitates supervision to the latent features.
\end{itemize}

\begin{table}
	\centering
	\caption{The MAP of the proposed MAH method equipped with the cascaded multi-head structure and deep hashing baselines with various code lengths of \{\textcolor{black}{12}, 24, 36, 48\}.}
	\resizebox{0.4\textwidth}{!}{%
		\begin{tabular}{l|cccc}
			\toprule
			\multirow{1}{*}{Method} &
			\multicolumn{3}{c}{CIFAR-10} &
			\\
			& {12 bits} &{24 bits} &{36 bits} &{48 bits}\\
			\midrule
			DPSH &65.34 &67.29 &70.13 &71.25\\
			\midrule
			DRSCH &67.54 &67.89  &68.32 &68.57\\
			\midrule
			ADSH &92.84 &94.21 &94.32 &93.75\\
			DAPH &75.69 &82.13 &83.07 &84.48 \\
			\midrule
			MAH-2 &\textbf{79.86} &\textbf{($4$-bit)} &\textbf{94.35} &\textbf{($8$-bit)} \\
			\bottomrule
		\end{tabular}%
	}
	\label{tab:2}
\end{table}


	\begin{figure*}[t]
		\centering
		\includegraphics[width=0.24\textwidth]{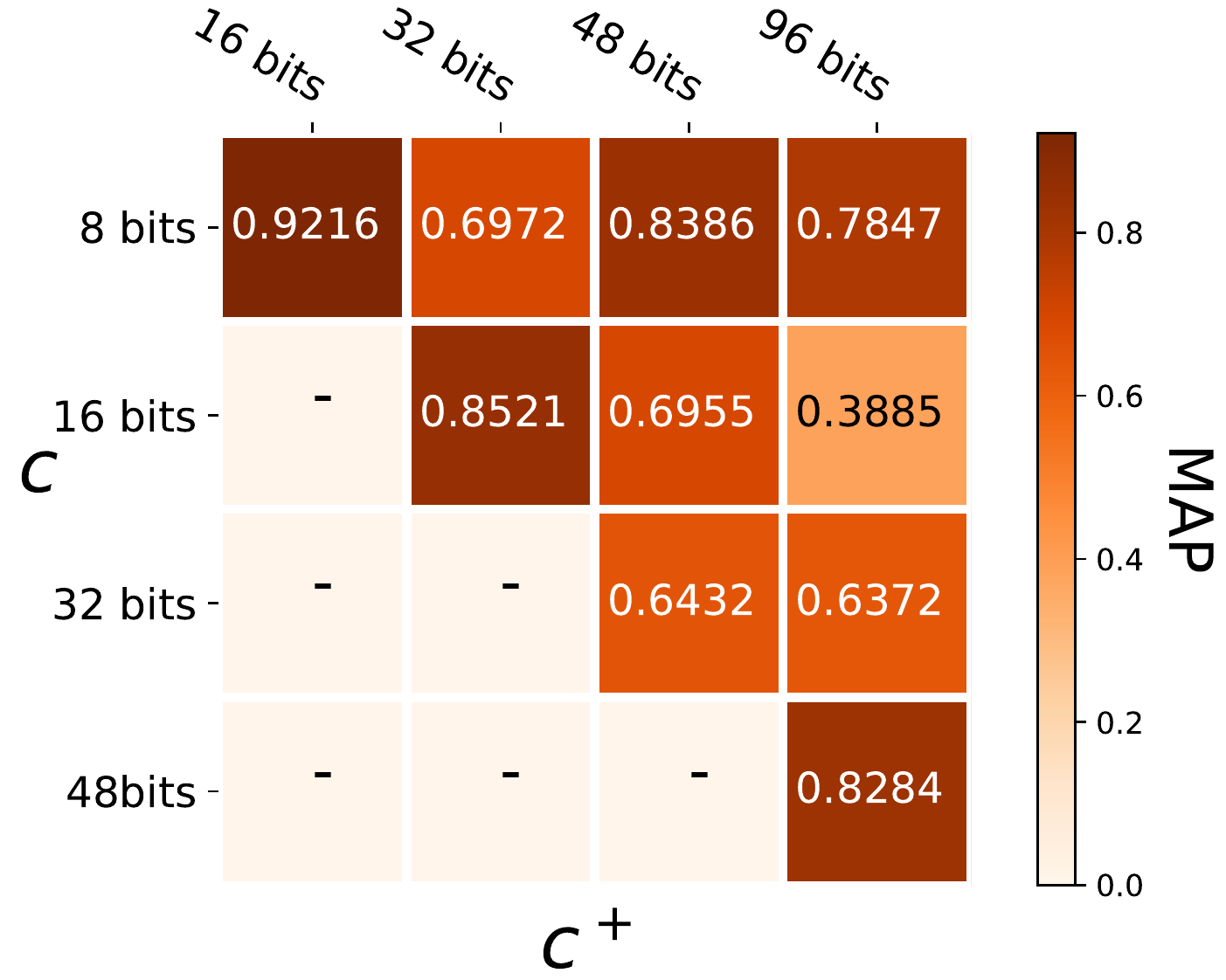}
		\includegraphics[width=0.24\textwidth]{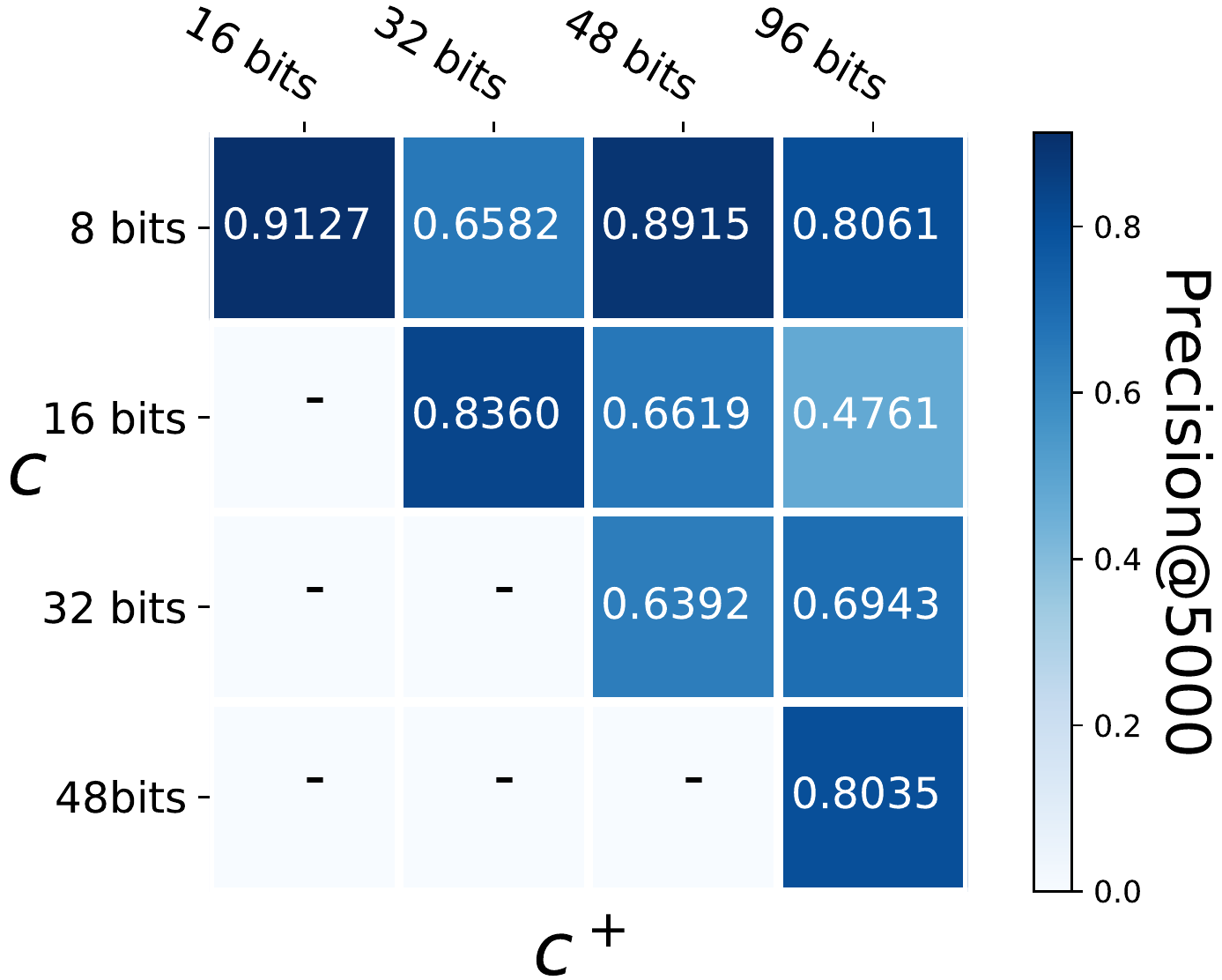}
		\includegraphics[width=0.24\textwidth]{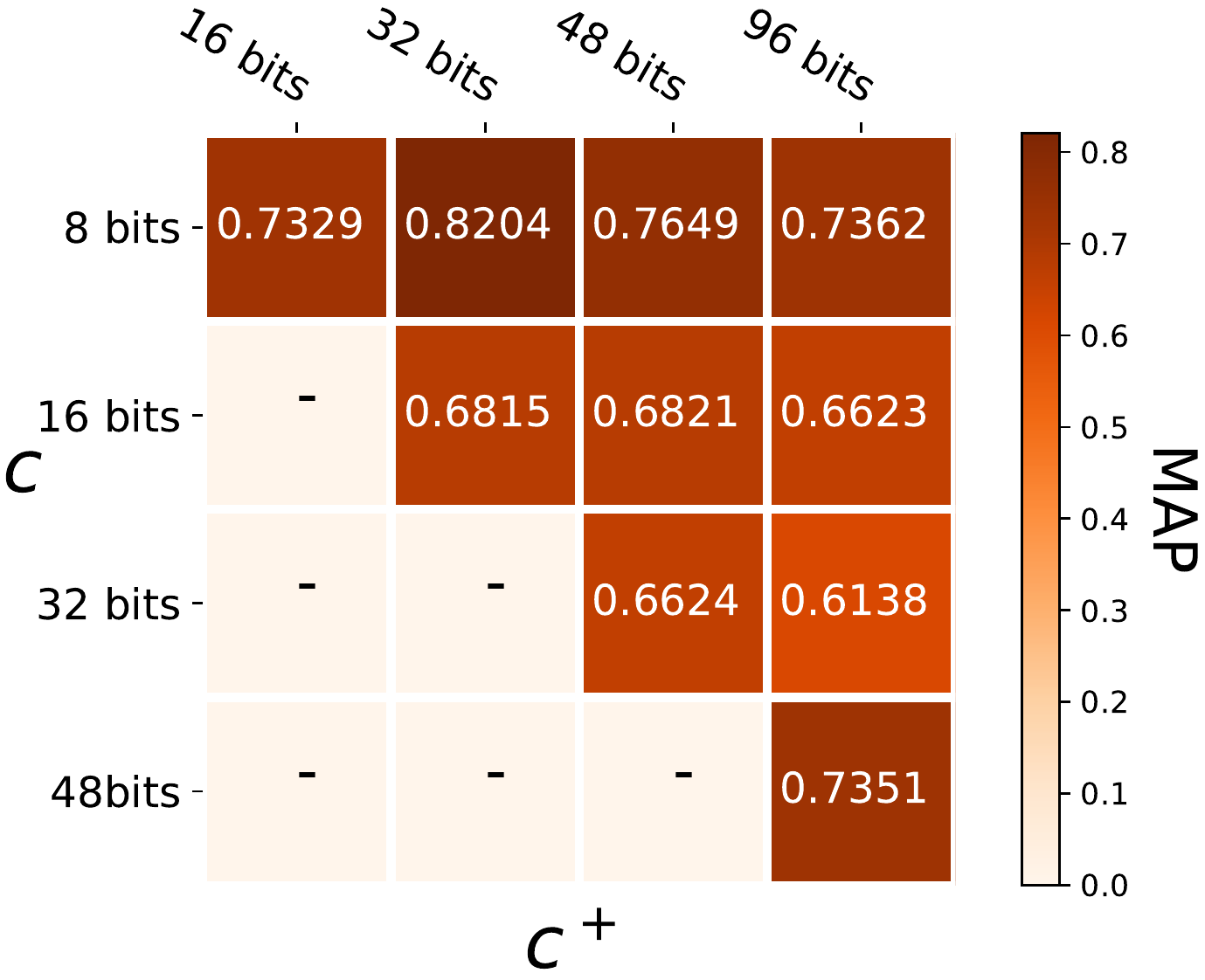}
		\includegraphics[width=0.24\textwidth]{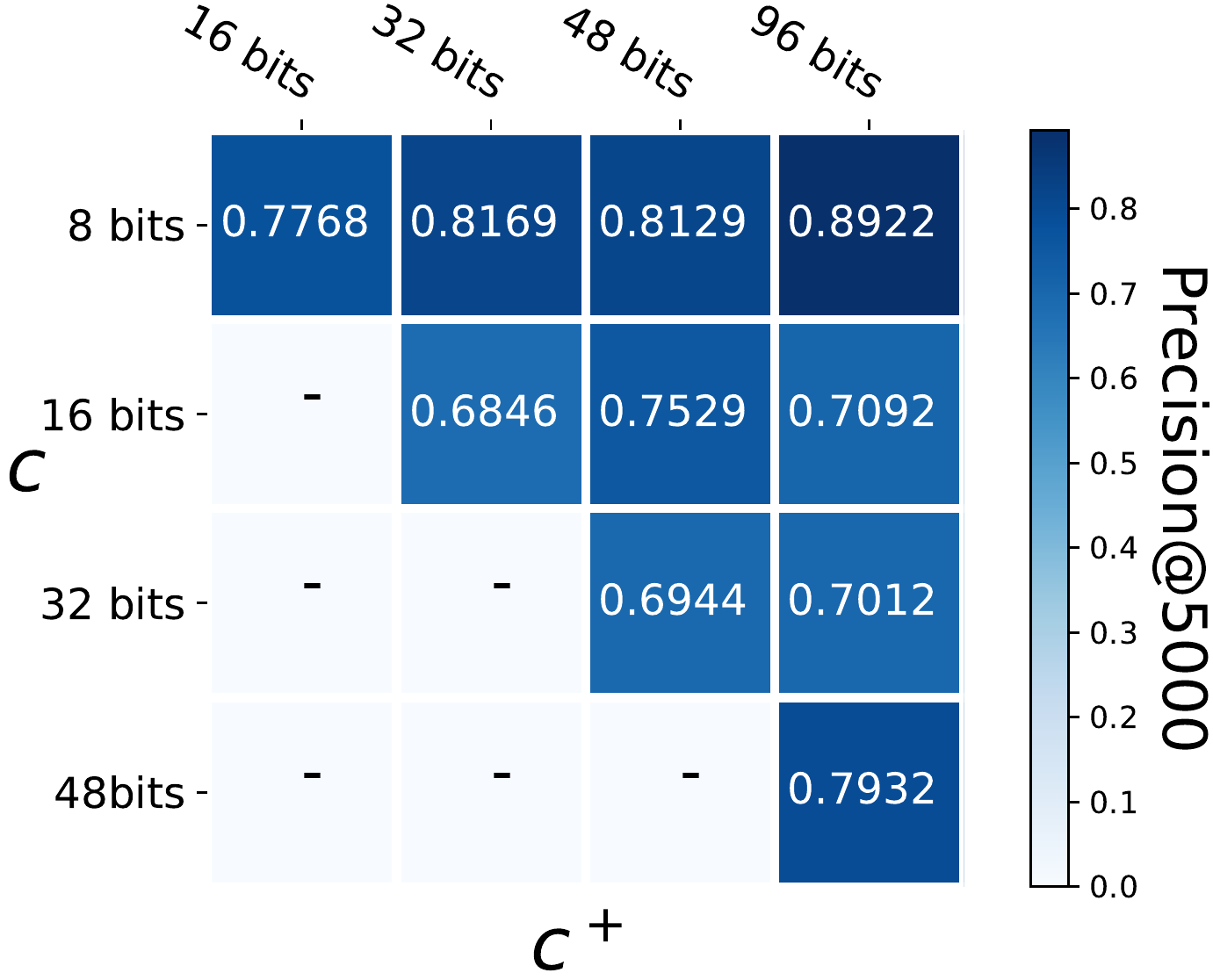}
		\caption{Ablation study on the large-scale multi-head strides ($8$, $16$, $32$, $48$, $96$). The performance of $c^-=6$ bit hash codes learned by MAH with the flat multi-head (\textit{Left}) and cascaded multi-head (\textit{Right}) on CIFAR-10 dataset. The loss coefficients  $\alpha$ and $\beta$ are fixed at 4 and 2 respectively.}
		\label{fig:distiller1}
	\end{figure*}

\begin{table*}[t]
	\caption{The ablation study of multi-head embedding branch given a fixed code length stride on the CIFAR-10 dataset.}
	\centering
	\label{tab:stride}
	\resizebox{0.9\textwidth}{!}{%
		\begin{tabular}{cccccc|cccccc}
			\toprule[0.04cm]	
			\multicolumn{12}{c}{Flat Multi-head (MAH-1)}\\
			\cmidrule{3-10}
			\multicolumn{6}{c}{MAP}&\multicolumn{6}{c}{Precision@5000} \\
			{4 bits} & {6 bits} & {8 bits} & {10 bits} & {12 bits}& {14 bits} &{4 bits} & {6 bits} & {8 bits} & {10 bits} & {12 bits}& {14 bits}\\
			\textbf{74.60} &88.88 &92.49 &- &- &- &\textbf{76.89} &87.39 &91.24 & - & - &- \\
			- &86.22  &94.29 &94.89 &- &-  & - &\textbf{91.31} &93.23 &93.80 &- &- \\
			- &- &\textbf{94.60} &94.30 &94.33 &-  &- &-&\textbf{93.65} &93.12 &93.20 &- \\
			- &- &- &\textbf{94.93} &94.83 &94.86 &- &- &-&\textbf{93.82} &93.49 &93.89 \\
			\midrule[0.03cm]
			\multicolumn{12}{c}{Cascaded Multi-head (MAH-2)}\\
			\cmidrule{3-10}
			\multicolumn{6}{c}{MAP}&\multicolumn{6}{c}{Precision@5000} \\
			{4 bits} & {6 bits} & {8 bits} & {10 bits} & {12 bits}& {14 bits} &{4 bits} & {6 bits} & {8 bits} & {10 bits} & {12 bits}& {14 bits}\\
			\textbf{51.30} &81.06 &91.38 &- &- &- &\textbf{63.62} &83.77 &89.59 & - & - &-\\
			- &76.12  &92.90 &93.69 &- &-  & - &82.69 &91.90 &92.11 &- &- \\
			- &- &\textbf{93.42} &94.38 &94.50 &-  &- &-&\textbf{92.42} &93.15 &93.16 &- \\
			- &- &- &\textbf{94.77} &95.37 &95.34 &- &- &-&\textbf{93.71} &94.39 &94.05\\
			\bottomrule[0.04cm]	
		\end{tabular}%
	}
\end{table*}

\subsection{Ablation Study}
Regarding the impact of each component and parameter setting, we set up ablation studies on stride of multi-head structures, loss coefficients, and hyper-parameters respectively, on the CIFAR-10 dataset.

\subsubsection{Effect of the Multi-Head Structures}
In this subsection, we conduct the ablation study on the effect of the stride of multi-head structure. We explore the retrieval performance with various strides of our multi-head structures and report the MAP and Precision@5000 in Fig. \ref{fig:distiller} and  Fig. \ref{fig:distiller1}. The Fig. \ref{fig:distiller} shows a local relationship, i.e., $c^-=4$ and $\{c, c^+\} \in \{6, 8, 10, 12, 14\}$. Extra experiments in a macro view, i.e., $c^-=6$ and $\{c, c^+\} \in \{8, 16, 32, 48, 96\}$ are shown in Fig. \ref{fig:distiller}. The loss coefficient $\alpha$ and $\beta$ are fixed at $4$ and $2$. It is clearly observed that,
\begin{figure*}[htb!]
	\centering
	\includegraphics[width=0.25\textwidth]{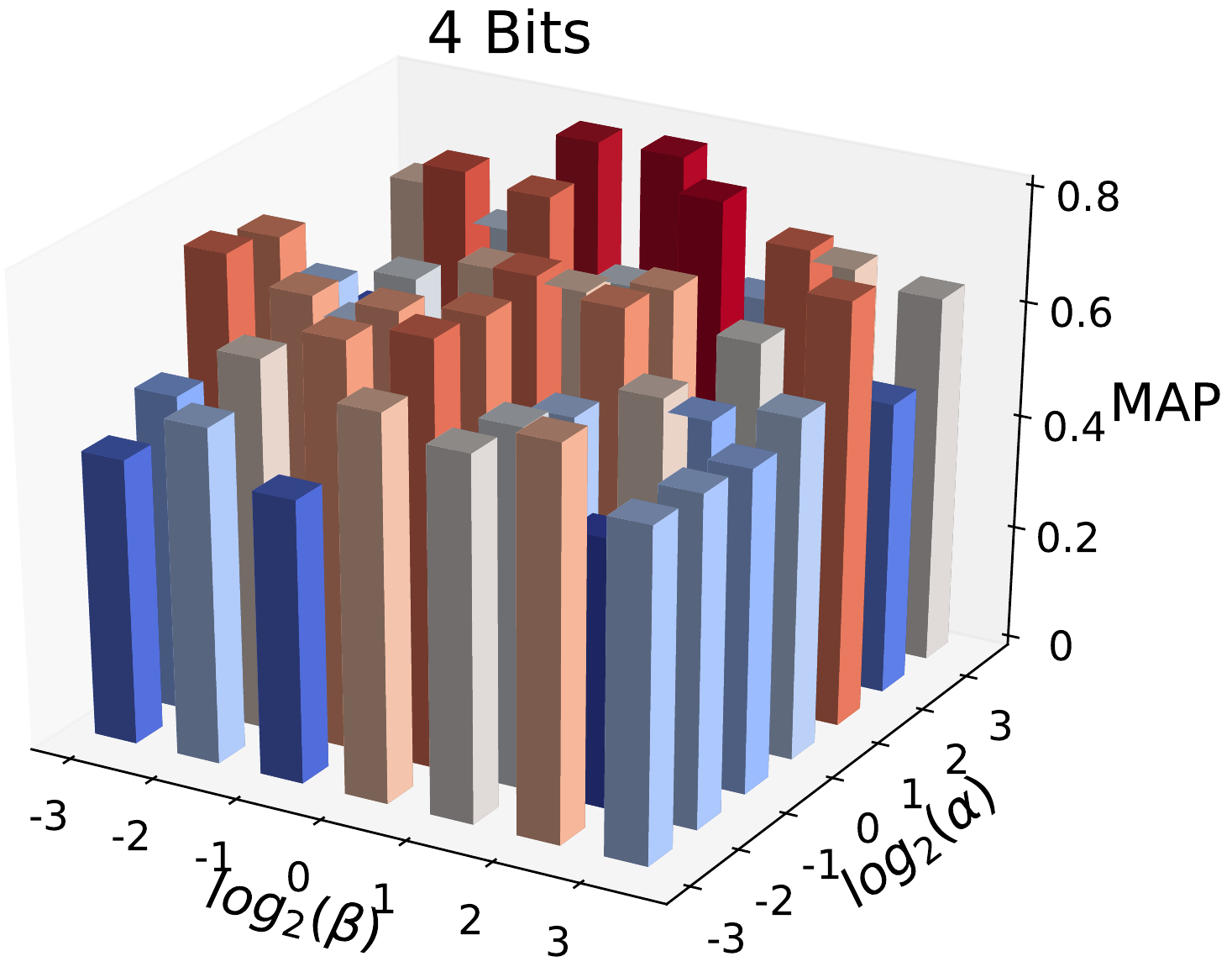}
	\includegraphics[width=0.25\textwidth]{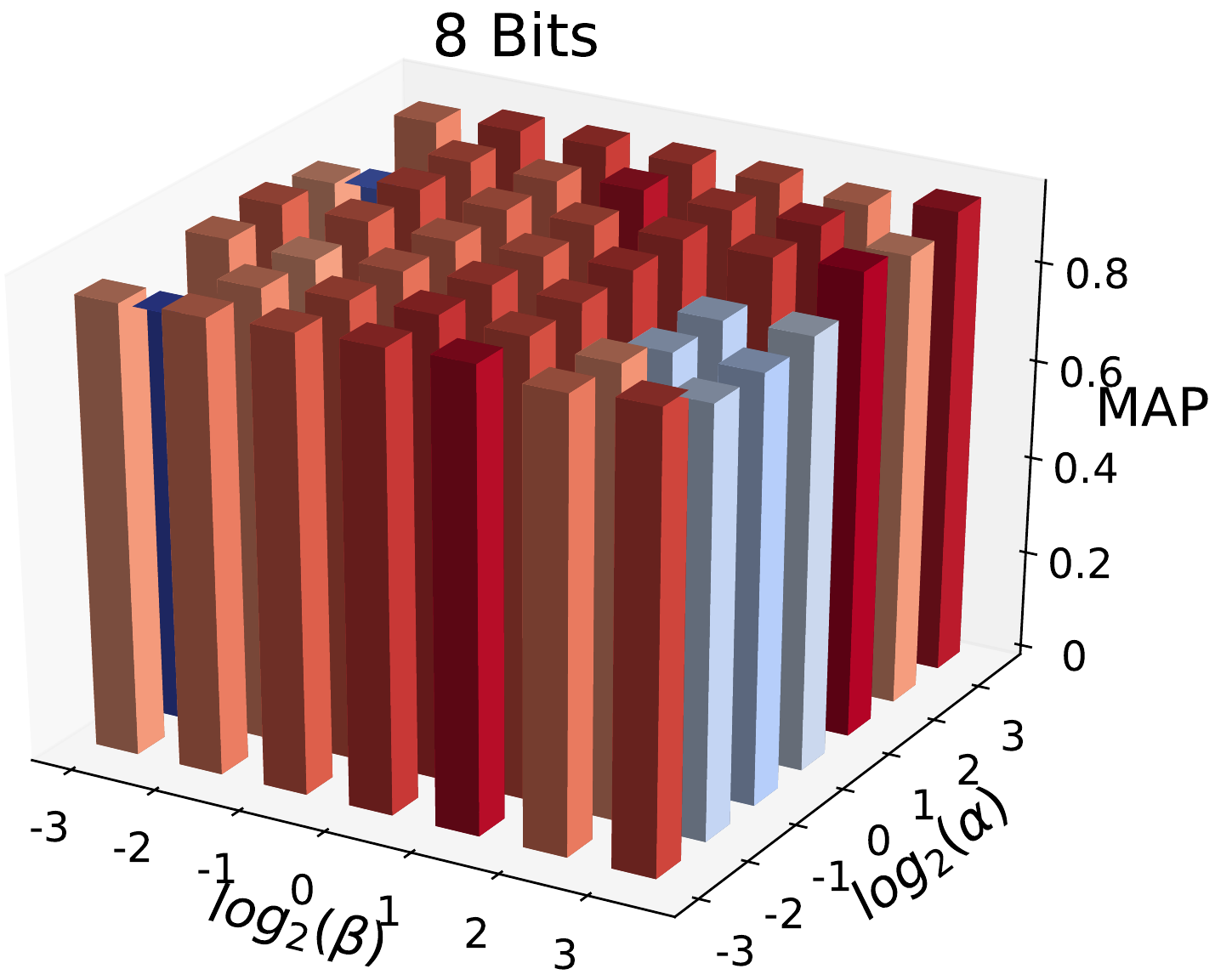}
	\includegraphics[width=0.25\textwidth]{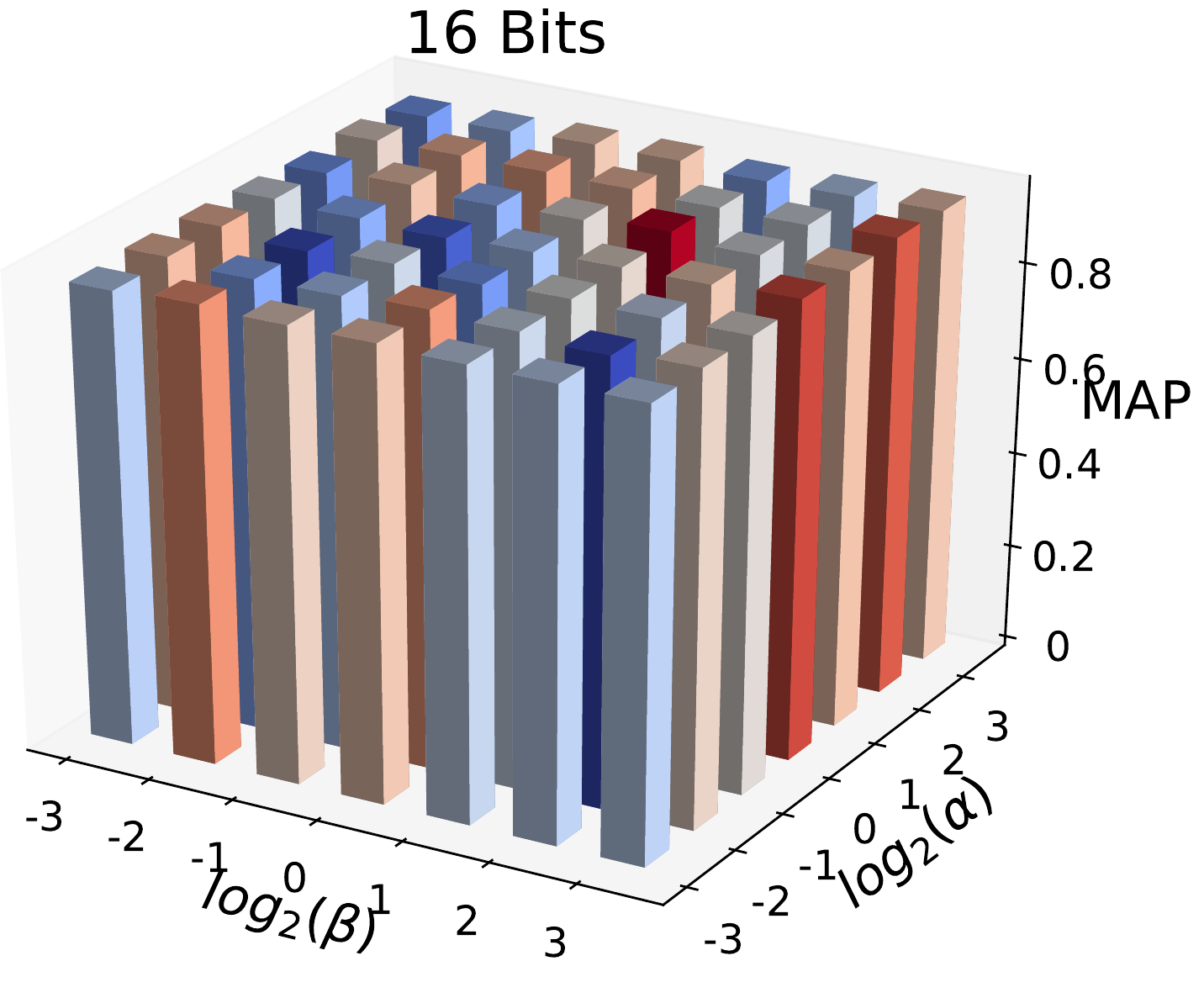}
	\caption{MAP of MAH-2 with various settings of loss coefficient $\alpha$ and $\beta$. (\textit{Left}) $c^-=4$, (\textit{Middle}) $c=8$, and (\textit{Right}) $c^+=16$.}
	\label{fig:coefficient_MAP}
\end{figure*}
\begin{figure*}[htb!]
	\centering
	\includegraphics[width=0.25\textwidth]{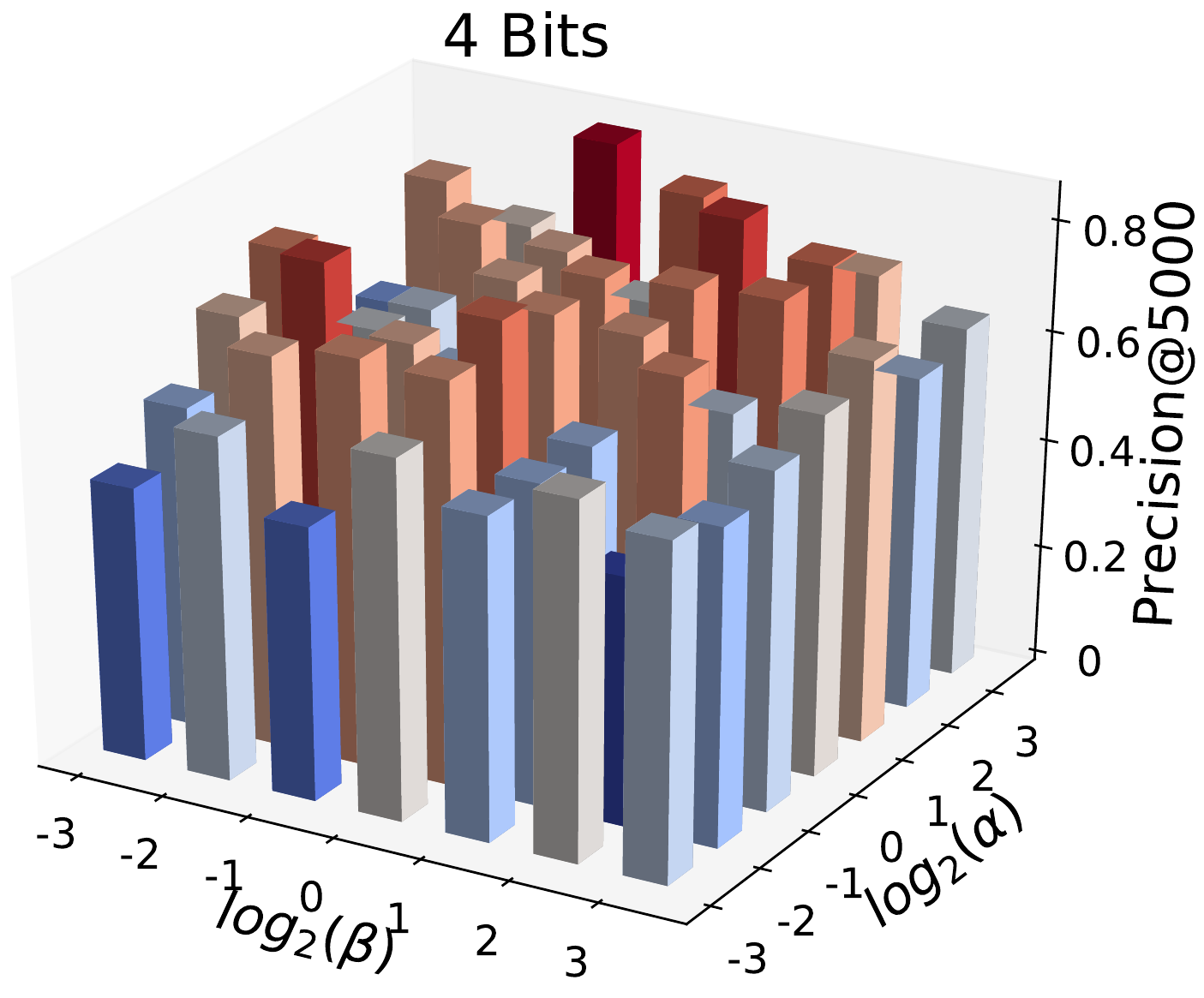}
	\includegraphics[width=0.25\textwidth]{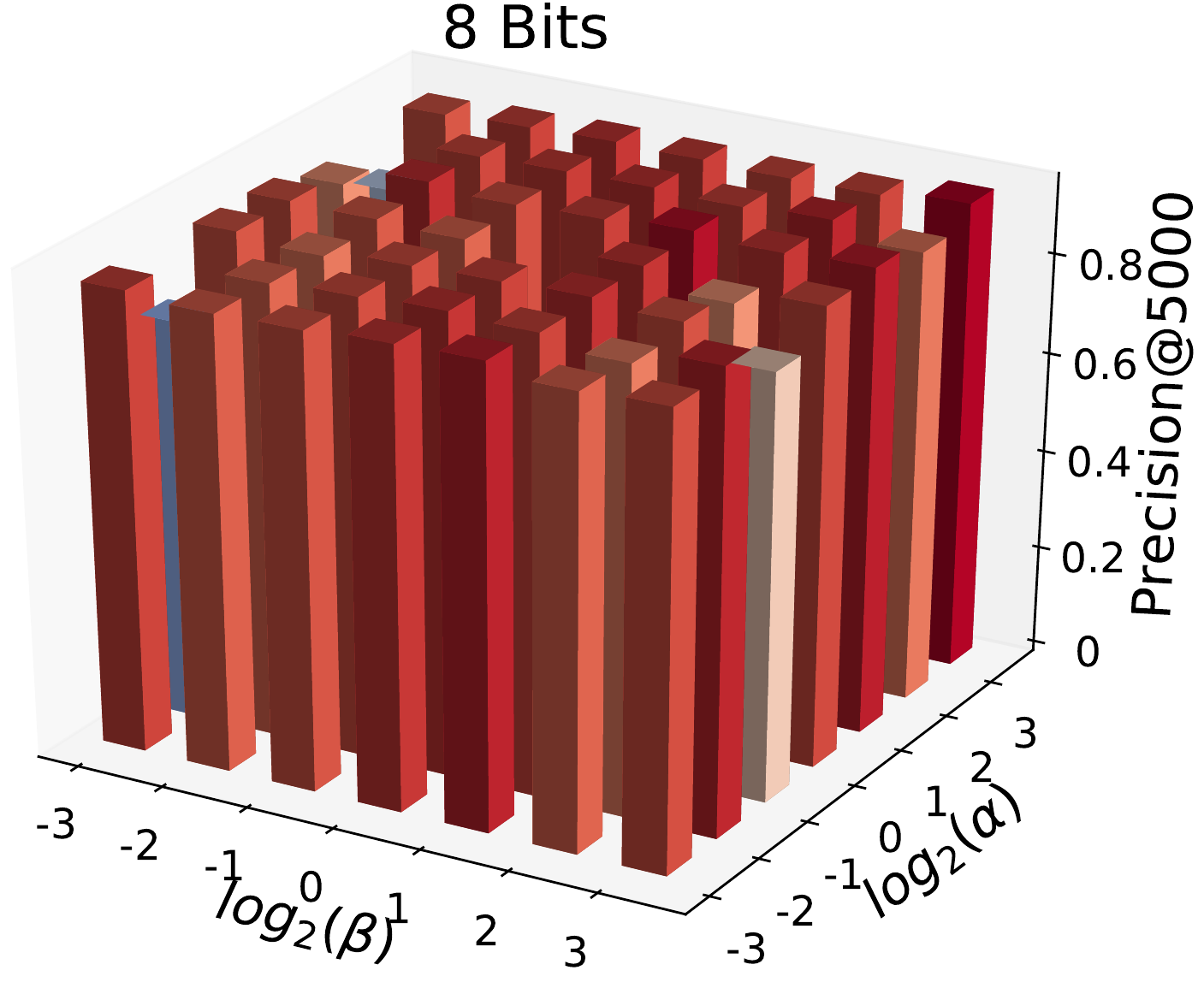}
	\includegraphics[width=0.25\textwidth]{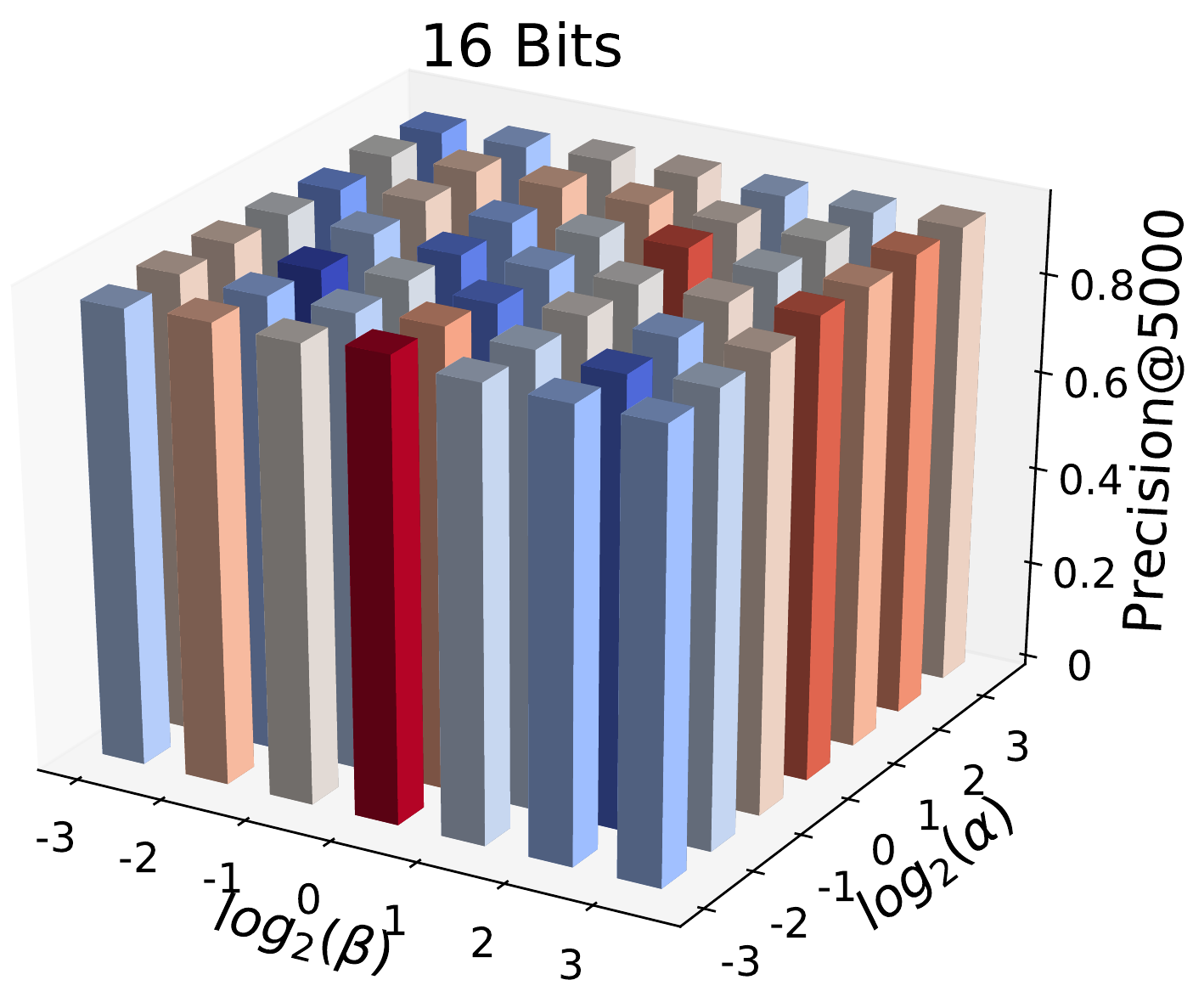}
	\caption{Precision@5000 of MAH-2 with various loss coefficient $\alpha$ and $\beta$. (\textit{Left}) $C^-=4$, (\textit{Middle}) $C=8$, and (\textit{Right}) $C^+=16$.}
	\label{fig:coefficient_Pre}
\end{figure*}
\begin{itemize}
	\item Given the fixed loss coefficients, the cascaded multi-head achieves a better score with the length combination of $\{4, 6, 8\}$ (74.60\% of MAP, 80.93\% of Precision@5000). In the horizontal and vertical directions, the indicator value gradually falls before slightly going up. We analyze this phenomenon results from the trade-off between decreasing the fitness of loss coefficients and increasing explicit knowledge from high-bit learning. More specifically, the learning of multiple hash codes will be dominated by a greater portion of high-bit learning when $c$ or $c^+$ is enlarged, which probably weakens the $c^-$-bit learning.
	\item Regarding the MAH with the flat multi-head, it peaks at $53.25\%$ of MAP and $72.01\%$ of Precision@5000, with the code length combination of $\{4, 8, 14\}$. The performance fluctuated more considerably in comparison to that with the cascaded multi-head.
\end{itemize}

Moreover, we investigate the impact of each embedding branch given a fixed code length stride, which is shown in Table \ref{tab:stride}.
From the Table \ref{tab:stride}, it is observed that, 
\begin{itemize}
	\item Given a target code length, the $c^-$-bit embedding branch achieves high scores in most of cases. As usually longer hash codes preserve more original data structure and semantics, they pass shorter hash codes with positive guidance and regularization.  
\end{itemize}
	As illustrated in Fig. \ref{fig:distiller1},
	\begin{itemize}
		\item The cascaded distiller reaches a higher score when $c:c^+ = 1:2$. For example, MAP and Top-5K are up to $92.16\%$ and $91.21\%$ respectively when $c=8$, $c^+=16$, of which the main reason we believe is the consistency with the loss coefficient $\beta=2$. The low bit learning could benefit from the normalized and balanced gradient of high-bit learning with different levels. 
		\item Dissimilar pattern is observed from MAP and Precision@5000 matrix of the flat distiller. It climbs to $83.04\%$ on MAP and $81.69\%$ on Precision@5000 precision, when $c^-=6, c=8, c^+=32$. Generally, it performs relatively well as $c=8$ in contrast to other cases, which may indicate that a small stride between $c^-$ and $c$ is required when applying the flat distiller into the end-to-end training. 
	\end{itemize}
To draw a conclusion, an excessive large stride of multi-head structures will not bring about a boost on retrieval performance, as its enlarged loss will dominate and overshadow the learning of low bit embedding. \eat{ Besides, as the length of hash codes goes up, not only the explicit knowledge but also the subsequent noise and redundancy might jeopardize the binary embedding.  }

\subsubsection{Effect of Loss Coefficients}
\eat{\eat{In this subsection, we mainly investigate the impact of the loss coefficients $\alpha$ and $\beta$ on the retrieval performance. }MAP and Precision@5000 are reported in Fig. \ref{fig:coefficient_MAP} and \ref{fig:coefficient_Pre}. We fix the code length $\{C^-, C, C^+\}$ to $\{4, 8, 16\}$. The ablation study is conducted on CIFAR-10 dataset with the MAH equipped with the cascaded distiller. 
	What we could find in Fig. $\ref{fig:coefficient_MAP}$ and Fig. \ref{fig:coefficient_Pre} is,}

The impact of the loss coefficients on embedding quality is reported in Fig. \ref{fig:coefficient_MAP} and Fig. \ref{fig:coefficient_Pre}. We fix the code length $\{c^-, c, c^+\}$ to $\{4, 8, 16\}$. The ablation study is conducted on CIFAR-10 dataset with the MAH equipped with the cascaded multi-head. 
It is observed from Fig. \ref{fig:coefficient_MAP} and Fig. \ref{fig:coefficient_Pre} that,
\begin{itemize}
	\item MAP of 4-bit hash codes rises to $79.86\%$ when $\alpha=4$ and $\beta=2$. Recall that $\{c^-, c, c^+\}$ are fixed to $\{4, 8, 16\}$, which is reverse to the ratio of loss coefficients $\{\alpha, \beta, 1\} = \{4, 2, 1\}$. Please note that our learned $4$-bit codes outperform most of the other $12$-bit codes learned by deep hashing methods or even $48$-bit codes (See Table \ref{tab:2}). 
	\item Most of high scores achieved in last three column when $\alpha \in\{2,4,8\}$, which shows weighting low bit embedding head gives a positive impact on its retrieval performance. 
	\item For auxiliary tasks, i.e, $8$- and $16$-bit learning, they have positive correlations with the low bit embedding, reaching $94.35\%$, $95.66\%$ of MAP and $93.88\%$, $94.70\%$ of Precision@5000.
\end{itemize}
To conclude, the setting of the loss coefficients directly influences the quality of the learned embeddings. The ratio of $\{\alpha, \beta, 1\}$ is suggested to be $\{c^+\pm\delta, c, c^-\}$ for the better performance, where $\delta$ is applied as a minor adjustment.
\eat{\subsubsection{Sensitivity to parameters}}

\subsection{The Study of Efficiency}
The proposed MAH algorithm is further studied with regard to training time and storage cost. 

\subsubsection{Training Efficiency}
In terms of training efficiency, we explore the training time of learning 4-bit hash codes by the proposed \textbf{MAH} and other deep hashing baselines \textbf{ADSH}, \textbf{DAPH}, \textbf{DPSH} and \textbf{DRSCH} on the CIFAR-10. The quantitative results are plotted in Fig. \ref{fig:training_map}. From  Fig. \ref{fig:training_map}, it is clearly observed that,
\begin{figure}[htb!]
	\centering
	\includegraphics[width=0.232\textwidth]{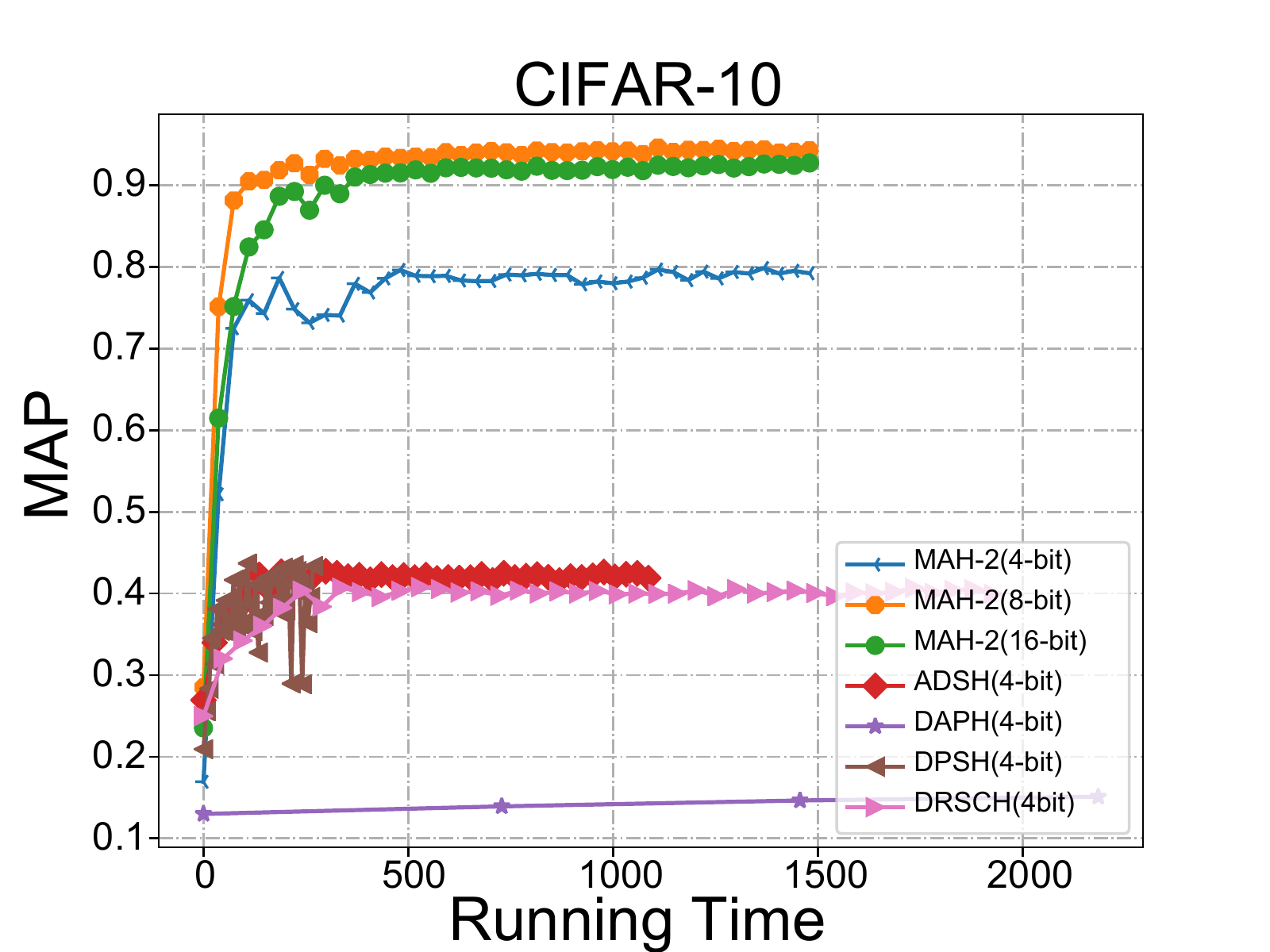}
	\includegraphics[width=0.232\textwidth]{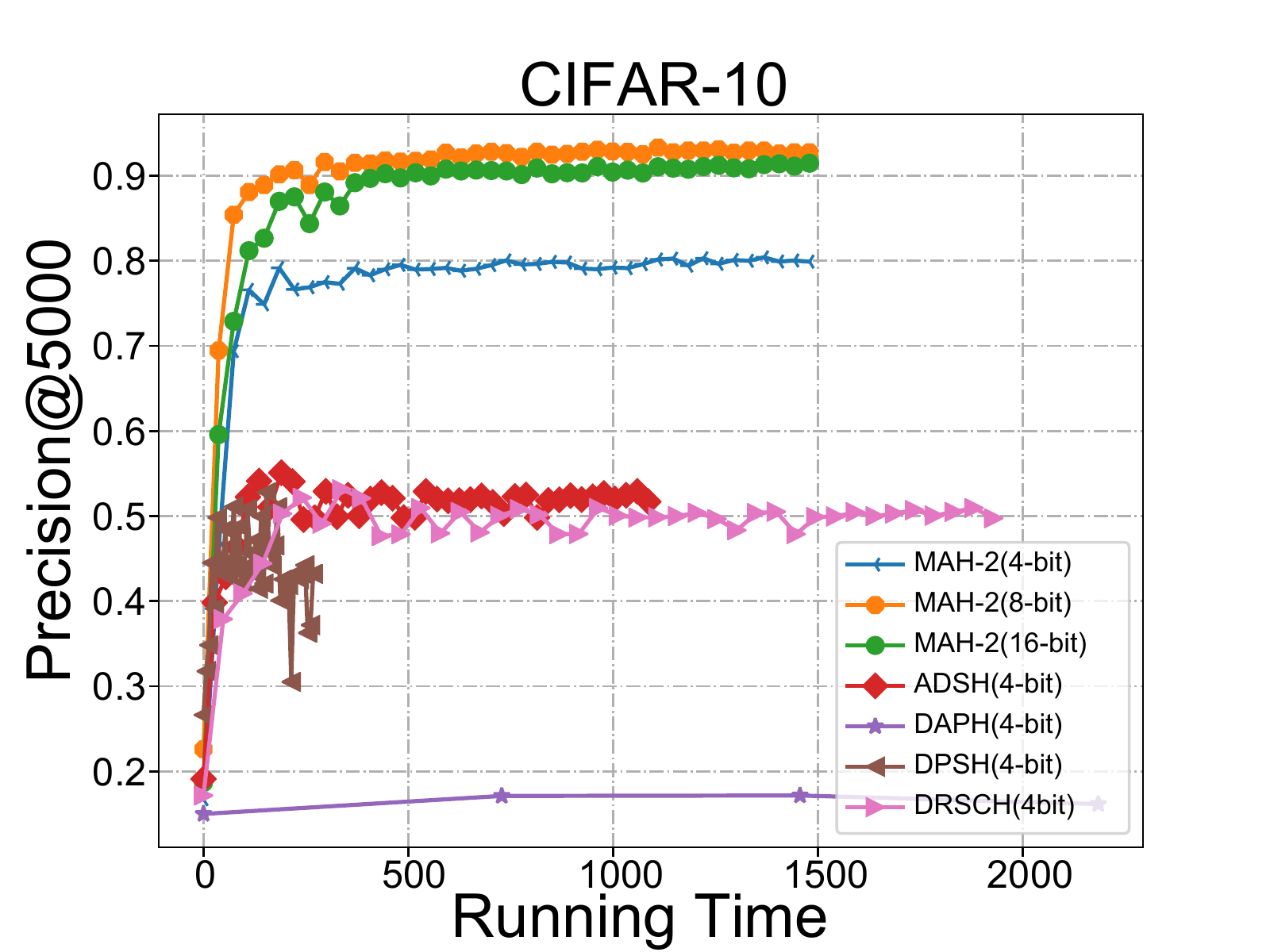}
	\caption{The training time and MAP of deep supervised hashing on the CIFAR-10 dataset.}
	\label{fig:training_map}
\end{figure}
\begin{itemize}
	\item The \textbf{MAH} achieves the first convergence at around the $6^{th}$ epoch. Both the MAP and Precision@5000 are superior to the state-of-the-arts.
	\item Due to its simplicity of structure, \textbf{DPSH} averagely consumes the least time per epoch ($6.70s$), while \textbf{DAPH} consumes $27.16s$, \textbf{MAH} $36.75s$, \textbf{DRSCH} $48.24s$. Training DAPH costs $19.69\times$ longer time in comparison with MAH, leading to $728.38s$ for each epoch. Note that MAH simultaneously learns hash codes with 3 different lengths, which is feasible especially when applied into practice.
	\item DAPH holds a stable performance before suddenly climbing up at approximately the $19^{th}$ epoch, achieving $53.48\%$ on MAP, followed by fluctuations in a slight decreasing trend. 
\end{itemize}

	\begin{figure}[t]
		\centering
		\includegraphics[width=0.45\textwidth]{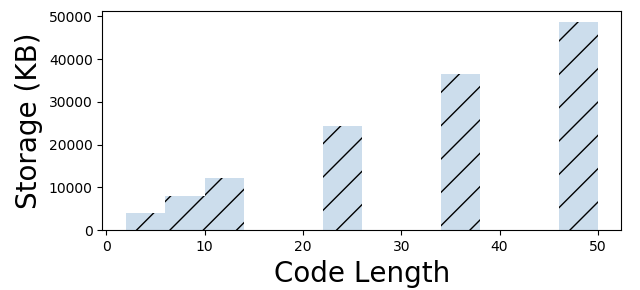}
		\caption{The cost of storing database hash codes for the CIFAR-10 dataset with various code lengths.}
		\label{fig:storage}
	\end{figure}

\subsubsection{Storage Cost}
With experimental results provided in Table \ref{tab:2}, the learned 4-bit hash codes by \textbf{MAH} achieve a better performance compared with the 48-bit binary codes learned by \textbf{DPSH} and \textbf{DRSCH}, and the 8-bit codes surpass the 48-bit codes by other deep hashing methods. Generally, the storage of database hash codes grows up linearly corresponding to code lengths, which is shown in Fig. \ref{fig:storage}. Therefore, we could infer that, by applying the proposed MAH into practice, the overall storage cost of hash codes will diminish by $83.33\%$ to $91.67\%$ without compromising on performance.

\section{Conclusion and Future Work}
In this paper, we propose the Multi-head Asymmetric Hashing (MAH) framework, pursuing the preservation of maximum semantic information with the minimum binary bits. By leveraging the flat and cascaded multi-head structures, the proposed MAH distills the bit-specific knowledge for low-bit codes with the guidance of other hashing learners and achieves a promising performance in low-bit retrieval tasks. Extensive experiments on three datasets have proven the superiority of the proposed MAH to the existing deep hashing methods, increasing MAP by $39.5\%$ and saving storage by $83.33\%$ to $91.67\%$. The proposed collaborative learning strategy is planned to extend on the neural network quantization task in a near future, where a significant compression and acceleration is expected.

\ifCLASSOPTIONcompsoc
  \section*{Acknowledgments}
\else
  \section*{Acknowledgment}
\fi

This work is partially supported by ARC FT130101530 and NSFC No. 61628206.

\ifCLASSOPTIONcaptionsoff
  \newpage
\fi



%
\balance{
\bibliographystyle{IEEEtran}
\bibliography{IEEEabrv,main}}

%

\begin{IEEEbiography}[{\includegraphics[width=1in,height=1.25in,clip,keepaspectratio]{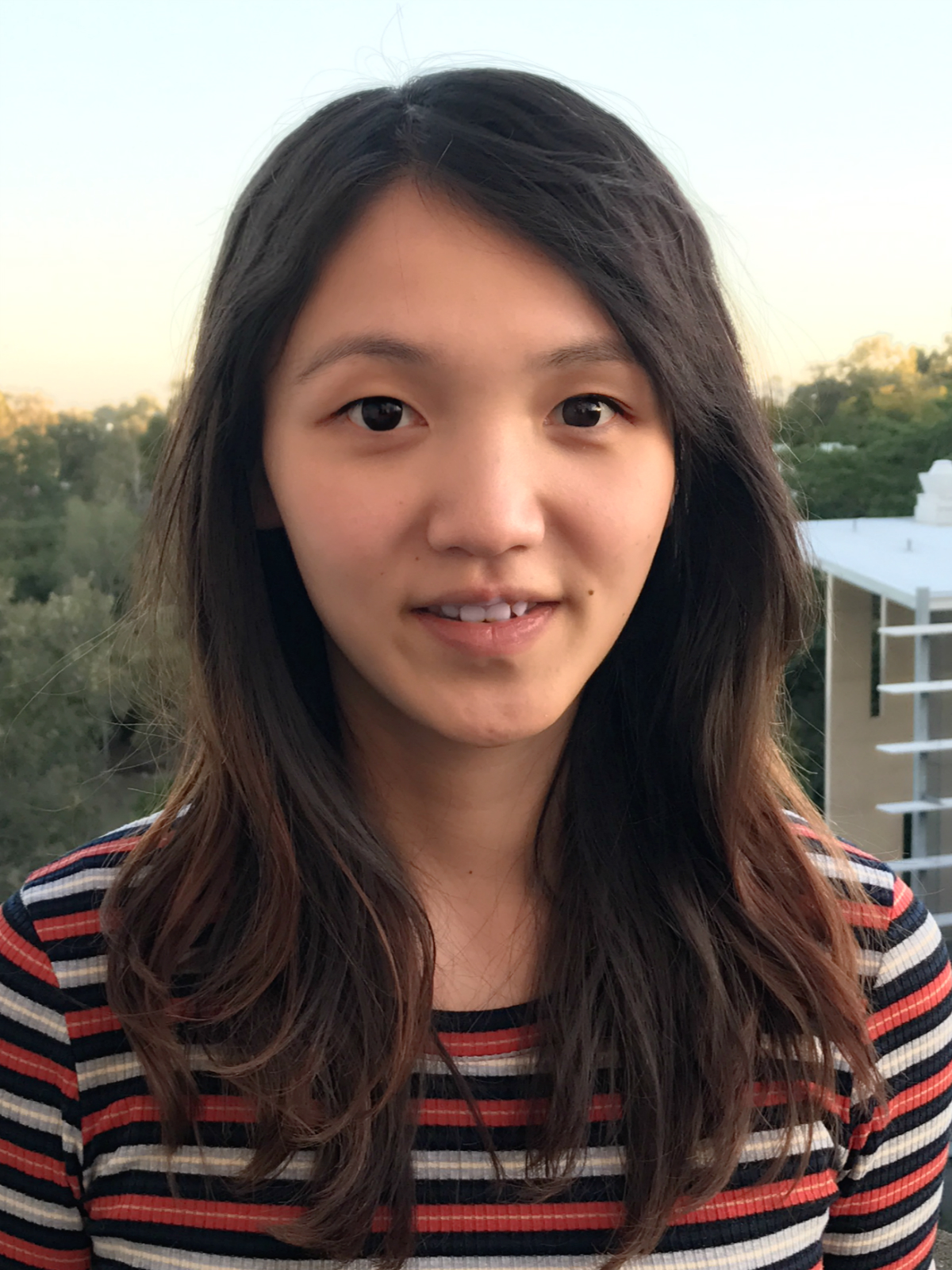}}]{Yadan Luo}
	received the B.S. degree in computer science from the University of Electronic Engineering and Technology of China in 2017. She is currently working toward the Ph.D. degree at the University of Queensland. Her research interests include multimedia retrieval, machine learning and computer vision.
\end{IEEEbiography}

\begin{IEEEbiography}[{\includegraphics[width=1in,height=1.25in,clip,keepaspectratio]{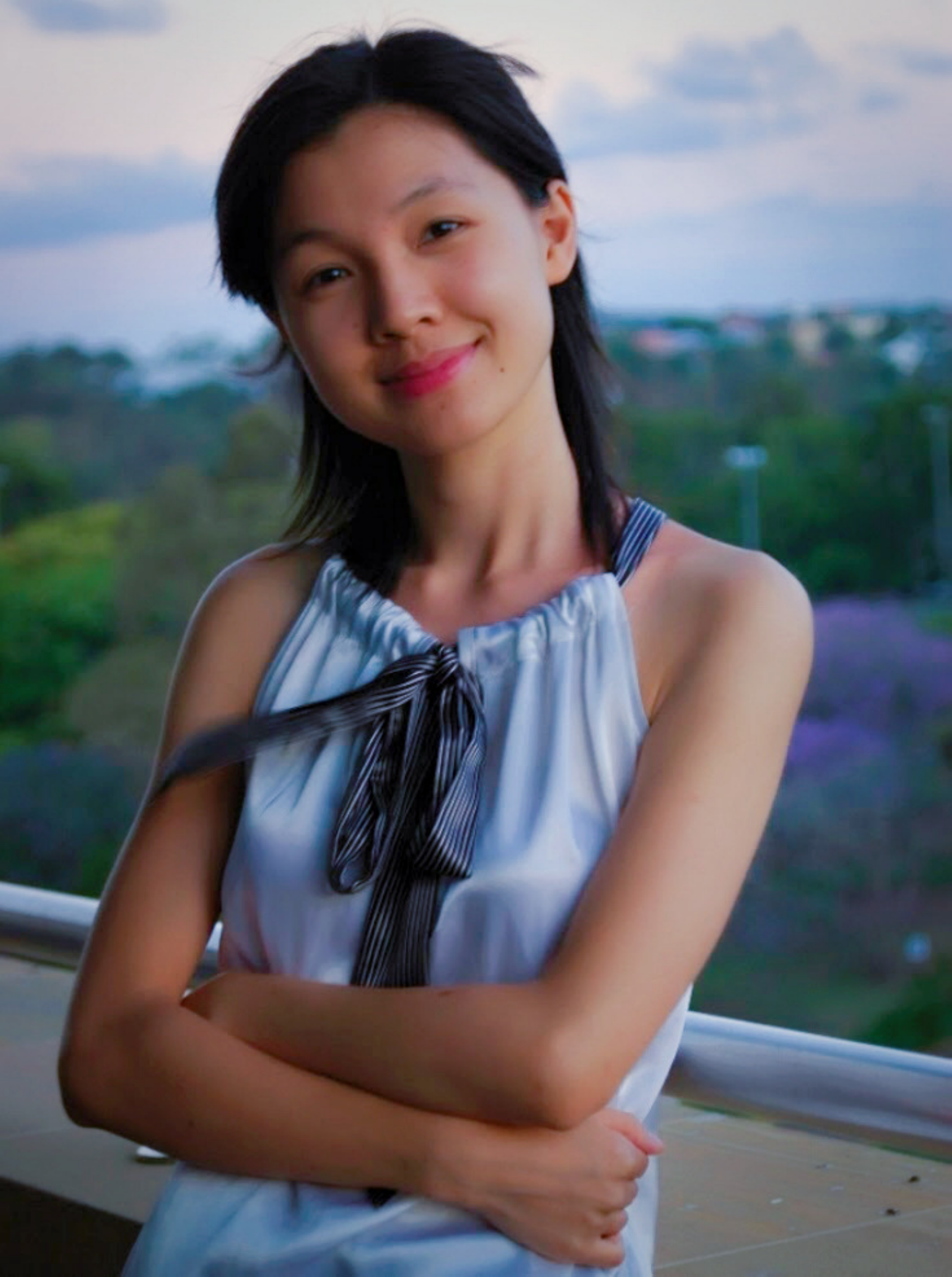}}]{Zi Huang}
	is an ARC Future Fellow in School of ITEE, The University
	of Queensland. She received her BSc degree from Department of Computer Science, Tsinghua University, China, and her PhD in Computer Science
	from School of ITEE, The University of Queensland. Dr. Huang's research
	interests mainly include multimedia indexing and search, social data analysis and knowledge discovery.
\end{IEEEbiography}

\begin{IEEEbiography}[{\includegraphics[width=1in,height=1.25in,clip,keepaspectratio]{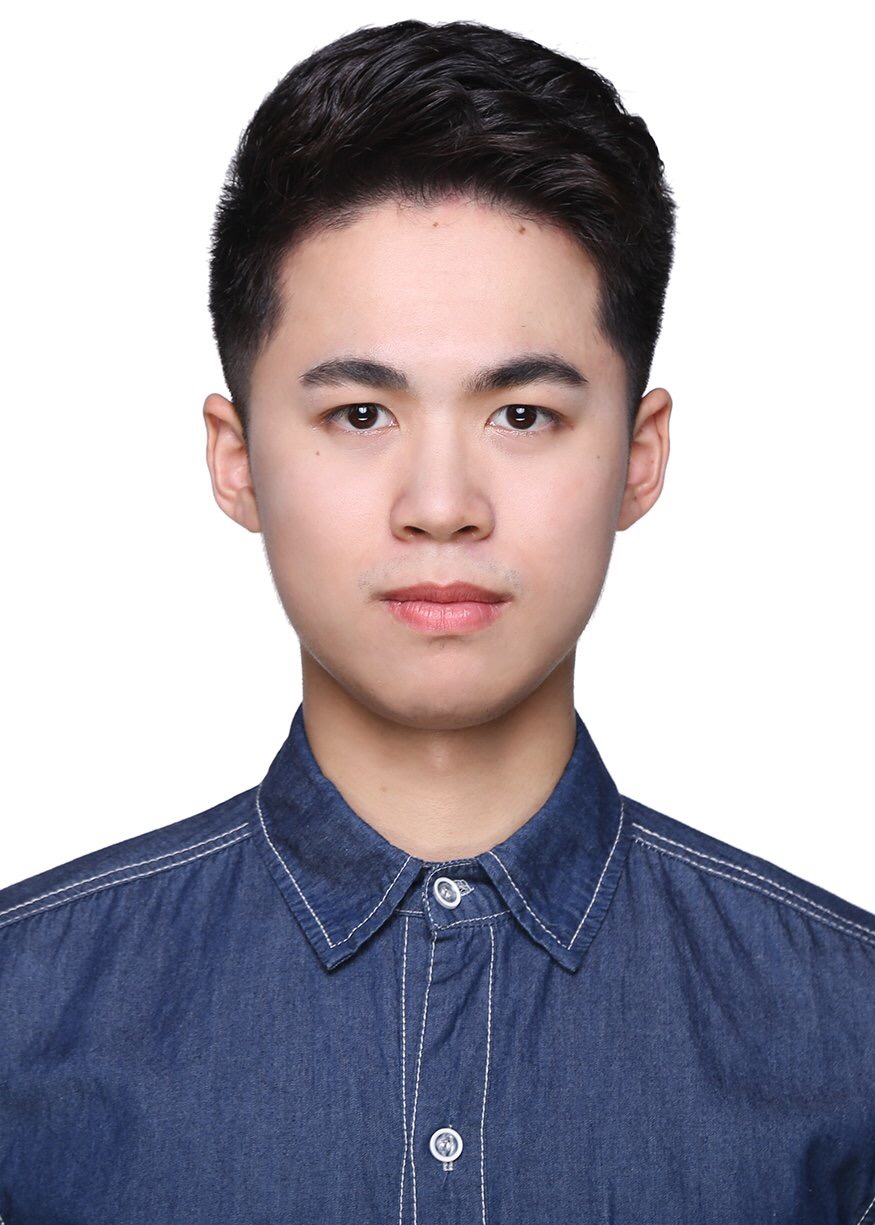}}]{Yang Li} received his BSc degree from Zhejiang Sci-Tech University in 2016 and his Master degree of Computer Science from The University of Queensland, Australia in 2018. He is currently a PhD candidate at The University of Queensland. His research interests include machine learning and recommendation. 
\end{IEEEbiography}

\begin{IEEEbiography}[{\includegraphics[width=1in,height=1.2in,clip,keepaspectratio]{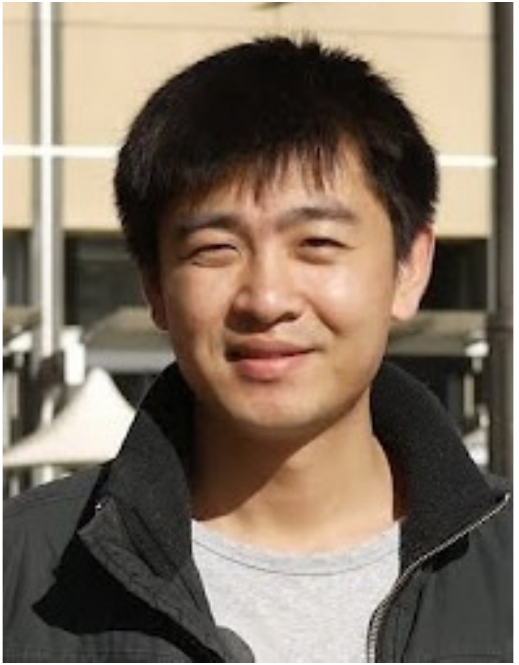}}]{Fumin Shen}
	received his Bachelor degree at 2007 and PhD degree at 2014 from Shandong University and Nanjing University of  Science and Technology, China, respectively. Now he is an associate professor of University of Electronic Science and Technology of China. His major research interests include computer vision and machine learning, including face recognition, image analysis and hashing methods.
\end{IEEEbiography}

\begin{IEEEbiography}[{\includegraphics[width=1in,height=1.25in,clip,keepaspectratio]{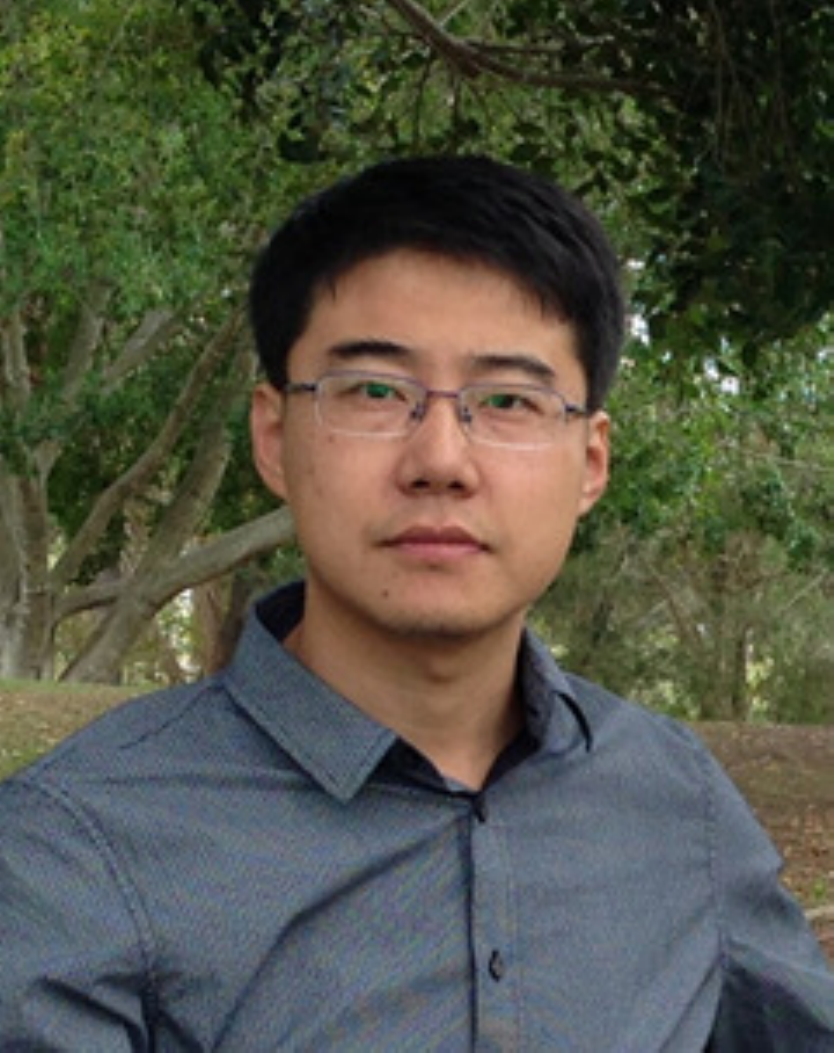}}]{Yang Yang} received the bachelor’s degree from
	Jilin University in 2006, the master’s degree from Peking University in 2009, and the Ph.D. degree from The University of Queensland, Australia,
	in 2012, under the supervision of Prof. H. T. Shen and Prof. X. Zhou. He was a Research Fellow under the supervision of Prof. T.-S. Chua with the National University of Singapore from 2012 to 2014. He is currently with the University of Electronic Science and Technology of China.
\end{IEEEbiography}

\begin{IEEEbiography}[{\includegraphics[width=1in,height=1.25in,clip,keepaspectratio]{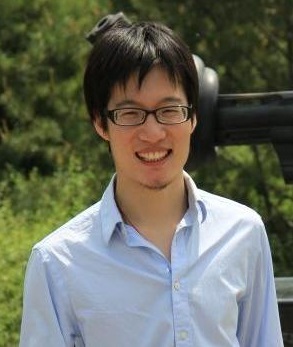}}]{Peng Cui} received his Ph.D. degree in computer science in 2010 from Tsinghua University and
he is an Associate Professor at Tsinghua. He has vast research interests in data mining, multimedia processing, and social network analysis. Until now, he has published more than 20 papers in conferences such as SIGIR, AAAI,
ICDM, etc. and journals such as IEEE TMM, IEEE TIP, DMKD, etc. Now his research is sponsored by National Science Foundation of China, Samsung, Tencent, etc. He also serves as Guest Editor, Co-Chair, PC member, and Reviewer of several high-level international conferences, workshops, and journals.
\end{IEEEbiography}




\end{document}